\documentclass[amsmath,amssymb,floatfix,nofootinbib,prd]{revtex4}
\pdfoutput=1
\usepackage{graphicx}
\usepackage{epsfig}
\usepackage{rotate}
\usepackage{amsmath}
\usepackage{amssymb}
\usepackage{amsfonts}
\usepackage{phoenician}
\usepackage{calc}
\usepackage{yfonts}

\usepackage{enumerate}
\usepackage{afterpage}
\usepackage{color}

\usepackage{soul}
\usepackage{ulem}

\usepackage{mathrsfs}

\usepackage{braket}

\usepackage{multirow}

\newcommand{\ud}{\mathrm{d}}
\newcommand{\p}{\partial}
\newcommand{\cH}{\mathcal{H}}
\newcommand{\Perp}{\mathcal{P}}

\newcommand{\D}{\mathcal{D}}

\newcommand{\T}{\mathcal{ T}}

\newcommand{\WW}{\mathbb{ W}}

\newcommand{\re}{{\rm e}}
\newcommand{\ro}{{\rm o}}

\def\be{\begin{equation}}
\def\ee{\end{equation}}
\def\bea{\begin{eqnarray}}
\def\eea{\end{eqnarray}}

\definecolor{darkred}{RGB}{141,3,20}

\newlength{\depthofsumsign}
\begin{document}

\begin{flushright}
CERN-TH-2019-153\\
KCL-PH-TH 2019-73
\end{flushright}


\title{Projection effects on the observed angular spectrum of the astrophysical\\
 stochastic gravitational wave background}

\author{Daniele Bertacca$^{a,b,c}$, Angelo Ricciardone$^{b}$, Nicola Bellomo$^{d,e}$, Alexander C. Jenkins$^{f}$, Sabino Matarrese$^{a,b,c,g}$, Alvise Raccanelli$^{h}$, Tania Regimbau$^{i}$, Mairi Sakellariadou$^{f}$}

\affiliation{
\vspace{0.25cm}
$^a$Dipartimento di Fisica Galileo Galilei, Universit\`{a} di Padova,  I-35131 Padova, Italy \\
$^b$INFN Sezione di Padova,  I-35131 Padova, Italy \\
$^c$INAF-Osservatorio Astronomico di Padova, Vicolo dell Osservatorio 5, I-35122 Padova, Italy\\
$^d$ICC, University of Barcelona, IEEC-UB, Mart\'i  i Franqu\`es, 1, E-08028 Barcelona, Spain.\\
$^e$Dept. de  F\'isica Qu\`antica i Astrof\'isica, Universitat de Barcelona, Mart\'i  i Franqu\`es 1, E-08028 Barcelona, Spain.\\
$^f$Theoretical Particle Physics and Cosmology Group, Physics Department, King's College London, University of London, Strand, London WC2R 2LS, UK\\
$^g$Gran Sasso Science Institute, Viale F. Crispi 7, I-67100 L'Aquila, Italy\\
$^h$Theoretical Physics Department, CERN, 1211 Geneva 23, Switzerland \\
$^i$ LAPP, Universit\'e Grenoble Alpes, USMB, CNRS/IN2P3, F-74000 Annecy, France\\
}

\begin{abstract}
The detection and characterization of the Stochastic Gravitational Wave Background (SGWB) is one of the main goals of Gravitational Wave (GW) experiments. 
The observed SGWB will be the combination of GWs from cosmological (as predicted by many models describing the physics of the early Universe) and astrophysical origins, which will arise from the superposition of GWs from unresolved sources whose signal is too faint to be detected.
Therefore, it is important to have a proper modeling of the astrophysical SGWB (ASGWB) in order to disentangle the two signals; moreover, this will provide additional information on astrophysical properties of compact objects.
Applying the {\sl Cosmic Rulers} formalism, we compute the  {\it observed} ASGWB angular power spectrum, hence using gauge invariant quantities, accounting for all effects intervening between the source and the observer. These are the so-called projection effects, which include Kaiser, Doppler and gravitational potentials effect.
Our results show that these projection effects are the most important at the largest scales, and they contribute to up to tens of percent of the angular power spectrum amplitude, with the Kaiser term being the largest at all scales.
While the exact impact of these results will depend on instrumental and astrophysical details, a precise theoretical modeling of the ASGWB will necessarily need to include all these projection effects.
\end{abstract}

\date{\today}

\maketitle
\bigskip
\section{Introduction}
The new run of observations from the LIGO/Virgo collaboration has recently started \cite{Aasi:2013wya} and many new Gravitational Wave (GW) from binary black hole (BBH), neutron star (BNS), and black hole-neutron star (BH-NS) mergers are being detected. One of the most challenging targets remains the detection (and characterization) of the background of gravitational waves (GWB). Such a background is generated by two contributions: a cosmological one originated from early universe-related mechanisms, and an astrophysical one, originated from the superposition of a large number of unresolved astrophysical sources.

Among the cosmological sources of GWs we can mention the irreducible GW background due to quantum vacuum fluctuations during inflation, which is expected to span over a wide range of frequencies, and for which we have already observational bounds from Planck~\cite{Akrami:2018odb}. In addition, inflation and post inflation-related mechanisms can generate a stochastic background of GWs at scales probed by interferometers like Laser Interferometer Gravitational-Wave Observatory (LIGO)/Virgo, LISA (Laser Interferometer Space Antenna) or ET (Einstein Telescope). For an overview of early universe GWB sources see \cite{Maggiore:1999vm, Guzzetti:2016mkm, Bartolo:2016ami, Caprini:2018mtu}.

On the astrophysical side, there are many sources that can contribute to form such a GW background (ASGWB), which is  the superposition of a large
number of unresolved sources and will be dominated by two types of events: the first is compact object binaries, periodic long lived sources such as early inspiraling phase of binary systems and captures by supermassive black holes, whose frequency is expected to evolve very slowly compared to the
observation time.
The second type consists of short-lived burst sources, such as core collapse to neutron stars or black holes, oscillation modes, r-mode instabilities in rotating neutron stars, magnetars and super-radiant instabilities (see \cite{Regimbau:2011rp, Romano:2016dpx} or for general reviews and references therein).

To characterize such backgrounds will be extremely challenging but necessary in order to extract precise cosmological information.
To a first approximation a cosmological background may considered stationary, isotropic, unpolarized and mainly Gaussian, while there are attempts to characterize
how a non-Gaussian and polarized background can be probed with interferometer like LISA or ground based interferometers \cite{Bartolo:2019oiq}.  The ASGWB has been usually characterized assuming that the distribution of sources is homogeneous and isotropic (and Gaussian).
The quantity which is commonly used to characterize the GWB, both of cosmological or astrophysical origin, is the GW energy density $\Omega_{\rm{GW}}$.
Beyond its isotropic value which has already invaluable information on the source of GWs, it can have a directional dependence inherited from the inhomogeneities of the matter distribution in the Universe, in a way similar to the Cosmic Microwave Background (CMB) radiation.
There has been a considerable effort in the GW community to detect such a background, but up to now we have only upper bounds on the isotropic GW energy density component  (LIGO/Virgo recent bounds are $\Omega_{\rm{GW}} (f = 25\,  \rm{Hz}) < 4.8 \times 10^{-8} $. Pulsar Timing Arrays (PTA), at low frequencies  ($10^{-10}  - 10^{-6} $ Hz),  gave  a bound  $\Omega_{\rm{GW}}< 1.3 \times 10^{-9}$  \cite{Shannon:2013wma}). Upper bound have been extracted also on its anisotropic component by LIGO and PTA (LIGO O1+O2 runs gave  $\Omega_{\rm{GW}} (f = 25\,  \rm{Hz}, ) < 6 \times 10^{-8} $ as upper limit~\cite{LIGOScientific:2019vic}   and PTA set   $\Omega_{\rm{GW}}  (f = 1 yr^{-1})< 3.4 \times 10^{-10} $ at 95\% CL~\cite{Taylor:2015udp}).
Such a background may be detectable with LIGO/Virgo at design sensitivity, especially with the addition of further interferometers to the global network (such as KAGRA and LIGO India).

In a series of recent works it has been shown how the anisotropy in the observed energy density of source distribution and the effect of inhomogeneities on the GW propagation can be used to infer astrophysical properties of the sources. A derivation of the angular power spectrum of  cosmological anisotropies, using a Boltzmann approach, has been obtained in \cite{Alba:2015cms, Contaldi:2016koz, Bartolo:2019oiq}.
 A derivation of the angular power spectrum of cosmological anisotropies, using a Boltzmann approach, has been obtained in \cite{Alba:2015cms, Contaldi:2016koz, Bartolo:2019oiq}. In the case of the ASGWB, the angular power spectrum has been derived by \cite{Cusin:2017fwz, Cusin:2018rsq, Cusin:2017mjm}, considering  the presence of inhomogeneities in the matter distribution and working with a coarse graining approach which allow to probe GW sources on cosmological, galactic and sub-galactic scales. Other predictions for the GW angular power spectrum have been derived in \cite{Jenkins:2018lvb, Jenkins:2018uac}, with both analytical and numerical results using galaxy catalogues from the Millennium simulation. More recently, \cite{Jenkins:2018kxc, Cusin:2019jpv, Cusin:2019jhg} have analyzed the astrophysical dependence of the angular power spectrum for different stellar models, while in~\cite{Jenkins:2019uzp, Jenkins:2019nks} the effect of shot noise on the angular power spectrum has been considered, and a new method to extract the true astrophysical spectrum by combining statistically-independent data segments has been proposed.
\vspace{0.03cm}

In this paper we present a consistent framework for studying the ASGWB in a general covariant setting. We  obtain general coordinate-independent and gauge-invariant results for all observables, accounting  for all effects intervening between the source and the observer. Working to linear order in perturbations, we investigate the effects  of cosmological perturbations and inhomogeneities on the angular power spectrum of the GW energy density.
Applying the {\sl Cosmic Rulers} formalism introduced in \cite{Jeong:2011as, Schmidt:2012ne}  (see also \cite{Bertacca:2017vod} where the authors used this prescription to study the effect of large-scale structures on GW waveforms),
we consider the observer's frame as the reference system.
In this case, all our results are obtained
at the observed frame, taking into account all
possible effects along the past GW-cone of the GW energy density.
It is important to note that the ASGWB is generated mostly by mostly events that could be in principle resolvable by 
precise and sensitive high resolution instruments. In principle we  might have a precise location of ASGWB in the observed space frame. Indeed,  the ASGWB signal is resembling other astrophysical backgrounds, such as e.g.,~the Cosmic Infrared Background (CIB), that have been studied in the past (see e.g.,~\cite{PlanckCIB, Tucci:2016hng, Lenz}).

Without any coarse graining but just mapping our perturbed quantities in the observer's frame, we obtain the corrections due to the inhomogeneous spacetime geometry. In a very general way, following  \cite{Cusin:2017fwz}, we consider two types of sources: (1) events with short emission, e.g., merging binary sources (BH-BH, NS-NS, NS-BH) and SNe explosions; (2) inspiraling binary sources which have not merged during a Hubble time.
We first work within a  general framework without fixing any gauge and we subsequently consider a $\Lambda$CDM concordance model on cosmological scales.
Using the perturbed GW energy density we then compute the {\it observed} angular power spectrum of the ASGWB highlighting the main local and integrated projection effects which give relevant contributions on large scales, considering a toy-model case: the ASGWB generated by black hole mergers in the frequency range of LIGO-Virgo. \vspace{0.1cm}

The paper is structured as follows.  In Section \ref{sec:formulation}, we define the GW energy density in a gauge independent way using the the observer's frame and we then give a general parametrisation for the description of the GW sources we will consider.
In Section \ref{sec:prescription}, we  study the past GW-cone in the observer's frame setting up the map between the observer's and real-space/physical frame and
 we then present a general perturbation framework of the quantities that enter in the GW energy density. In Section  \ref{sec:firstorder}, we do the computation of the perturbed quantities at linear level without fixing the gauge using a FLRW metric.  In Section \ref{sec:aps}, we focus on $\Lambda$CDM and compute the angular spectrum of the energy density using the Synchronous-Comoving gauge.   In Section \ref{sec:AngulaPS}, we compute the angular correlation between the energy density from different directions. Finally, in Section~\ref{sec:results}, we numerically evaluate the corrections for different contributions. We summarise  our conclusions in Section \ref{sec:conclusion}.  Through the text we will use $c=1$ and $(-,+,+,+)$ conventions.
 
\section{Covariant formulation of the GW energy density}
\label{sec:formulation}

The quantity that characterizes the SGWB is the GW energy density per logarithmic frequency  $f_\ro$, defined as~\cite{Allen:1997ad, Cusin:2017fwz}
\be
\Omega_{\rm GW}\left(f_{\rm o}, \Omega_o\right)= \frac{f_{\rm o}}{\rho_{\rm c}} \frac{\ud \rho_{\rm GW}}{\ud f_{\rm o} \ud \Omega_\ro } \,,
\ee
which represents the fractional contribution of gravitational waves to the critical energy density of the Universe today, $\rho_{\rm c} = 3H_0^2/(8\pi G)$, and $\rm {d} \rho_{\rm GW}$ the energy density of GWs   in the frequency interval $\{f,f +df\}$.
Such a quantity will have both a background (monopole) contribution in the observed frame, which is, by definition, homogeneous and isotropic (${\bar{\Omega}}_{\rm GW}/4 \pi$) \footnote{Since it is related to an angular average in the observed frame.}, and a direction-dependent contribution $\Omega_{\rm GW}(f_{\rm o}, \Omega_\ro  )$.
In this work we focus on the angular power spectrum of this second contribution (for other recent analyses, see~\cite{Cusin:2017fwz, Jenkins:2018lvb}).

The total gravitational energy density in a direction {\bf n} is the sum of the  all unresolved astrophysical \\contributions along the line of sight contained in a given volume $\ud V_\re ({\bf n})$
\be \label{rho_GW-1}
\frac{\ud \rho_{\rm GW}}{\ud f_{\rm o}  \ud \Omega_o}= \frac{\ud \mathcal{E}^{\rm tot}_{\rm GW}}{\ud f_o \ud \T_{\rm o} \ud A_{\rm o} \ud \Omega_{\rm o}} = \sum_{[i]}  \int  n_{\rm h}^{[i]}(x_\re^\alpha, \vec{\theta} )
\frac{\ud  {}{\mathcal{E}^{[i]}_{\rm GW}} (x^{\mu}_\re \to x^{\mu}_{\rm o}, \vec{\theta} )} {\ud f_{\rm o} \ud \T_{\rm o}  \ud A_{\rm o}}
 \left|\frac{\ud V_\re}{\ud \Omega_{\rm o} \ud \chi}\right| \ud\chi  \,  \ud\vec{\theta} \;,
\ee
where $[i]$ is the index of summation over all unresolved astrophysical sources that produce the background of GWs, $\vec{\theta}= \{M_{\rm h}, M^*, \vec{m}, \vec{\theta}^*\}$, where $M_{\rm h}$ is the halo mass,  $M^*$ is the  mass of  stars that give origin to the sources,  $\vec{m}$ are the masses of the compact objects  and $\theta^*$ includes the astrophysical parameters related to the model (i.e., spin, orbital parameters, star formation rate). Here, $n_{\rm h}^{[i]}$
 is the (physical)  number of halos at given mass $M_{\rm h}$, in the physical volume $\ud V_\re$, {\it weighted}  with the parameters $\vec{\theta}$ of the sources at $x_\re^\mu$.
 The letter ``${ \re}$'' stands for ``evaluated at the emission (source) position'' while ``${ \ro}$'' for ``evaluated at the observed position''.

The physical volume  $\ud V_\re$ at emission is defined as
\be
\ud V_e \equiv  \sqrt{-g(x^\alpha)}  \varepsilon_{\mu \nu \rho\sigma}
u^\mu(x^\alpha) \frac{\partial x^f}{\partial \bar x^1} \frac{\partial x^\rho}{\partial \bar x^2} \frac{\partial x^\sigma}{\partial \bar x^3}
 \ud^3\bar {\bf x} = {\mathcal D}_{\rm A}^2(z) (- u_\mu p_{\rm GW}^\mu)\ud \Omega_{\rm o} \ud \lambda= {\mathcal D}_{\rm A}^2(z) (- u_\mu p_{\rm GW}^\mu)\left| \frac{\ud \lambda }{ \ud \chi}\right|  \ud \Omega_{\rm o} \ud \chi\,,
\ee
where $\varepsilon_{\mu \nu \rho \sigma}$ is the Levi-Civita tensor,  $u_{\mu}$ is the four velocity vector as a function of comoving location, and we have introduced the angular diameter distance ${\mathcal D}_{\rm A}$
and the GW four-momentum $p_{\rm GW}^\mu$. Let us point out that,  as in \cite{Bertacca:2017vod}, here we consider {\it the local wave zone} approximation to define the tetrads at source position (i.e. the observer ``at the emitted position" is a region with a comoving distance to the source sufficiently large so that the gravitational field is ``weak enough" but still ``local'', i.e., its wavelength is small w.r.t. the comoving distance from the observer $\chi$ (see for example~\cite{Maggiore:1900zz}).

The four-velocity of the observer can be written using the comoving tetrad
\begin{eqnarray}
\label{u}
u^\mu = \frac{\ud x^\mu}{\ud \T}=\frac{\ud x^{\hat{\alpha}}}{\ud \T}\Lambda^\mu_{\hat{\alpha}}=u^{\hat{\alpha}}\Lambda^\mu_{\hat{\alpha}}= \Lambda^\mu_{\hat{0}}\;,
\end{eqnarray}
where $\T$ is the proper time of the observer 
 and $\Lambda^\mu_{\hat \alpha}$ is an orthonormal tetrad.
 Choosing $u^\mu$ as the time-like basis vector,
\begin{equation}
\label{E0mu}
u_\mu= \Lambda_{\hat{0} \mu}=a \, E_{\hat{0} \mu} \quad \quad {\rm and}  \quad \quad u^\mu=\Lambda^\mu_{\hat 0}= a^{-1}E_{\hat{0}}^\mu\;,
\end{equation}
where $E^{\hat \alpha}_\mu$ are the components of the comoving tetrad which are defined through the following  relations
\begin{eqnarray}
\label{LambdaE}
\hat g^{\mu \nu} E^{\hat \alpha}_\mu E^{\hat \beta}_\nu& = &\eta^{\hat  \alpha \hat \beta}\;, \quad  \quad  \eta_{\hat  \alpha \hat \beta} E^{\hat \alpha}_\mu E^{\hat \beta}_\nu = \hat g_{\mu \nu}\;,  \quad  \quad \hat g^{\mu \nu} E^{\hat \beta}_\nu = E^{\hat \beta \mu}\;,    \quad  \quad \eta_{\hat  \alpha \hat \beta}  E^{\hat \beta}_\nu  =  E_{\hat \beta \nu}  \;,
\end{eqnarray}
and $ \eta_{\hat  \alpha \hat \beta} $ is the Minkowski metric.  The graviton four-vector is defined as
\be \label{p_GW}
p_{\rm GW}^\mu = \frac{\ud \chi}{ \ud \lambda} {\ud x^\mu\over \ud \chi}=- {2\pi f_\ro \over a^2} k^\mu\;,
\ee
where $k^\mu$ is the comoving null four-vector of the GW, $u^\mu$ is the four-velocity of the observed at $x_\re^\mu$ and $\lambda$ is the affine parameter
can be written  (normalised) in the following way\footnote{This suitable normalisation in Eq.\ \eqref{dlambda} can be completely understood in Section \ref{sec:prescription}.}
\be \label{dlambda}
\ud \lambda =- {a^2\over 2\pi  f_\ro} \ud \chi\,,
\ee
from the source to the detector. Here $\chi$ is the comoving distance, in real-space, from the observer to the source of the GW. Clearly,  $(- u_\mu p_{\rm GW}^\mu )=2\pi f_\re$.

The energy of gravitational waves emitted from the halo at the observer is
\be \label{energy_flux}
\frac{\ud  {}{{\mathcal{E}^{[i]}_{\rm GW}}} (x^{\mu}_\re \to x^{\mu}_\ro,  \vec{\theta} ) } {\ud  f_\ro \ud \T_\ro  \ud A_\ro}= \frac{{\mathcal K}^{[i]}(z, f_\re, x^{\mu}_\re, \vec{\theta} )}{(1+z)^3 {\mathcal D}_{\rm A}^2(z)}\,,
\ee
where
${\mathcal K}^{[i]}(z, f_\re, x^{\mu}_\re)$ encodes all physical effects of the GW signal emitted where the superscript $[i]$ is related to a typical unresolved astrophysical source considered.

The quantity ${\mathcal{E}^{[i]}_{\rm GW}} (x^{\mu}_\re \to x^{\mu}_\ro)$ (to simplify the notation, we will write it  as ${\mathcal{E}^{[i]}_{\rm GW}}_\ro$), for given type of source labelled by $[i]$, can be related to the energy spectrum per unit solid angle in the rest frame of the observer (in the halo) that includes all emitting sources at a given redshift $z$ and direction ${\bf n}$, as
\bea \label{energy_flux-2}
\frac{\ud  {\mathcal{E}^{[i]}_{\rm GW}}_\ro } {\ud  f_\ro \ud \T_\ro  \ud A_\ro} = {\ud  {\mathcal{E}^{[i]}_{\rm GW}}_\ro  \over \ud  {\mathcal{E}^{[i]}_{\rm GW}}_\re} ~
{\ud  f_\re \ud \T_\re \over \ud  f_\ro \ud \T_\ro }~  {\ud \Omega_\re \over \ud A_\ro}~ \frac{\ud  {\mathcal{E}^{[i]}_{\rm GW}}_\re } {\ud  f_\re \ud \T_\re  \ud \Omega_\re}\;,
\eea
where $\ud  {\mathcal{E}^{[i]}_{\rm GW}}_\re /\ud  f_\re  \ud \Omega_\re$ is the energy spectrum per unit solid angle of observer with $z \equiv z_\re$.

Using the energy conservation
\be
{\ud  {\mathcal{E}^{[i]}_{\rm GW}}_o  \over \ud  {\mathcal{E}^{[i]}_{\rm GW}}_\re} = {1 \over (1+z)},
\ee
and the relations
$ {\ud  f_\ro \ud \T_\ro = \ud  f_\re \ud \T_\re }$, and
\be
{\ud \Omega_\re \over \ud A_\ro} = {1 \over (1+z)^2 {\mathcal D}_{\rm A}^2(z)}\,,
\ee
we can rewrite Eq.(\ref{energy_flux-2}) as
\be
\label{Spectrum}
\frac{\ud  {\mathcal{E}^{[i]}_{\rm GW}}_\ro } {\ud  f_\ro \ud \T_\ro  \ud A_\ro} ={1 \over (1+z)^3 {\mathcal D}_{\rm A}^2(z)}  \frac{\ud  {\mathcal{E}^{[i]}_{\rm GW}}_\re } {\ud  f_\re \ud \T_\re  \ud \Omega_\re}\;.
\ee
In general, for a particular type of source, $\ud  {\mathcal{E}^{[i]}_{\rm GW}}_\re /\ud  f_\re /\ud \T_\re  /\ud \Omega_\re$ has a specific distribution function characterised by  local parameters of the source which depends on the mass, environment, distribution of matter, velocity dispersion of the matter and source, and the type of galaxies within the host halo. We can thus distinguish two cases: (I) events with short emission ({\it{burst sources}}), e.g. merging binary sources (BH-BH, NS-NS and/or NS-BH) and SNe explosions; (II) inspiraling binary sources which have not merged during a Hubble time, and hence GW emission is averaged over several periods of the slow evolution of the orbitals parameters ({\it{continuous sources}}).
The resulting energy in the two cases reads
\begin{eqnarray}
\label{}
\frac{\ud  {\mathcal{E}^{[i]}_{\rm GW}}_\re } {\ud  f_\re \ud \T_\re  \ud \Omega_\re} =  \;
\left\{
 \begin{array}{ll}
  {\ud {{ \mathcal N}^{[i]}_{\rm GW}}_\re  \over   \ud \T_\re }      \frac{\ud  {\mathcal{E}^{[i]}_{\rm GW}}_\re } {\ud  f_\re \ud \Omega_\re}
   & {\rm for}~~ (\rm I)\; \\ \\
{{\mathcal N}^{[i]}_{\rm GW}}_\re {\ud A_\re\over \ud \Omega_\re}  \frac{\ud  {\mathcal{E}^{[i]}_{\rm GW}}_\re } {\ud  f_\re \ud \T_\re  \ud A_\re }
& {\rm for} ~~ (\rm II)\;,
\end{array} \right.
\end{eqnarray}
where  for case (I), $\ud {{ \mathcal N}^{[i]}_{\rm GW}}_\re /   \ud \T_\re $ is the merging rate of the events for each halo and $\ud  {\mathcal{E}^{[i]}_{\rm GW}}_\re  / \ud  f_\re/ \ud \Omega_\re$ is the energy spectrum per unit solid angle, while for case (II)
\be \label{Spectrum-2}
\frac{\ud  {\mathcal{E}^{[i]}_{\rm GW}}_\re } {\ud  f_\re \ud \T_\re  \ud A_\re } = \overline{
 {\langle \tau_{\rm GW}^{\hat 0 \hat 0}}^{[i]} \rangle}_\re= \frac{1}{16 \pi G} \,   \overline{\left\langle \sum_{A=(+,\times)}  f_\re^2 \,{{\mathcal A}_\re^2}_A\right\rangle}.
\ee
Here ${\mathcal A}_\re$ is the amplitude at emission\footnote{We are  using {\it the local wave zone} approximation, hence the coordinates are strictly related on the considered halo.} and we have decomposed  the above quantity in the two independent modes of linear polarisation of the GWs.
The overline in Eq. (\ref{Spectrum-2}), denotes the ``time average" of the observer.
Following  \cite{Arnowitt:1961zz}, we have defined $\langle ... \rangle$ as the average over a region whose characteristic dimension is small compared to the scale over which the background changes.
The average is over the emitted region whose characteristic dimension is  about the scale of the halo dimension (around $1-2$Mpc).
We can thus identify the quantity ${\mathcal K}^{[i]}(z, f_e, x^{\mu}_\re)$ as the energy at emission
\be
\frac{\ud {\mathcal{E}^{[i]}_{\rm GW}}_\re } {\ud  f_\re \ud \T_\re  \ud \Omega_\re} = {\mathcal K}^{[i]}(z, f_\re, x^{\mu}_\re,  \vec{\theta} )\,
\ee
and obtain the following  expression for  the energy density
\be \label{rho_GW-2}
\frac{\ud \rho_{\rm GW}}{\ud  f_\ro \ud \Omega_\ro} = \sum_{[i]} \int  a(x^0)^2 {n^{[i]}(x_\re^\alpha,  \vec{\theta} )  \over (1+z)^2} \ud \chi  \ud\vec{\theta} \;,
\ee
where we define the {\it total} GW density as
\be
n^{[i]}(x_\re^\alpha, \vec{\theta} )\equiv n_h^{[i]}(x_\re^\alpha) \frac{\ud {\mathcal{E}^{[i]}_{\rm GW}}_\re } {\ud  f_\re \ud \T_\re  \ud \Omega_\re}(z, f_\re, x^{\mu}_\re, \vec{\theta} ) \;.
\ee

\section{General prescription}
\label{sec:prescription}
Let us define $x^\mu (\chi)$ the comoving coordinates in the {\it real frame} (or real space, the ``physical frame"), where $\chi$ is the comoving distance, in real-space, from the source to the detector (the observer) and call {\it observer's} the frame where we perform observations; we will adopt the approach of \cite{Bertacca:2017vod}. Assuming the usual concordance background model, let us use coordinates which effectively flatten our past gravitational wave-cone so that the GW geodesic from the source has conformal space-time coordinates:
\begin{equation}
\label{Obframe}
\bar{x}^\mu=(\bar \eta,\; \bar {\bf x})=(\eta_0-\bar \chi, \; \bar \chi \, {\bf n}).
\end{equation}
Here $\eta_0$ is the conformal time at observation, $\bar \chi(z)$ is the comoving distance to the observed redshift, ${\bf n}$ is the observed direction of the GW, i.e. $n^i=\bar x^i/\bar \chi=\delta^{ij} (\p \bar \chi/\p \bar x^j)$. Using $\bar \chi$ as an affine parameter in the observer's frame, the total derivative along the past GW-cone is $\ud / \ud \bar \chi = - \p/ \p \bar \eta + n^i \p/\p \bar x^i$.
We use again subscripts  ``{\re}" and ``{\ro}" to denote respectively the position where the GW is emitted and received.
The frame  defined in Eq. (\ref{Obframe}) is the real observed frame in which we make observations (also called ``cosmic GW laboratory'' in \cite{Bertacca:2017vod}). Therefore, this is the correct frame where, for instance, we can reconstruct 3D maps/catalogs of  galaxies  by using  both EM and GW signals. This frame is commonly used in galaxy catalogs. 
If we use unperturbed coordinates we are not able to interpret correctly the correlation between the ASGWB and EM sources from  observed galaxies since it can induce a bias in our results. 

Defining the photon 4-momentum $p_{\rm GW}^\mu=-2\pi f_\ro k^\mu/a^2$, where $a$ is the scale factor,
the comoving null geodesic vector $k^\mu$ reads
 \begin{equation}
\label{kmu}
 k^\mu(\bar \chi) = \frac{\ud  x^\mu }{\ud \bar \chi}(\bar \chi)= \frac{\ud }{\ud \bar \chi}  \left(\bar{x}^\mu + \delta x^\mu\right)(\bar \chi) = \left(-1+\delta  f,\; n^i+\delta n^{i } \right)(\bar \chi)\;,
\end{equation}
with
\begin{equation}
\label{kmu-0}
\bar{k}^\mu=\frac{\ud  \bar{x}^\mu }{\ud \bar \chi}=\left(-1, \; {\bf n} \right)\;.
\end{equation}
The comoving coordinate in the physical frame can be written as
\be
\label{xphi}
x^\mu ( \chi) = \bar{x}^\mu (\chi)+ \delta x^{\mu } (\chi)= \bar{x}^\mu (\bar \chi)+ \frac{\ud  \bar{x}^\mu }{\ud \bar \chi} \delta \chi+
                                     \delta x^{\mu} (\bar \chi),
\ee
with
\begin{eqnarray}
\label{chi}
\chi = \bar \chi+ \delta \chi \, ,
\end{eqnarray}
\be \label{delta-x}
\delta x^{\mu}=\delta x_\ro^{\mu} + \int_0^{\bar \chi_\re}  \delta k^\mu \; \ud\bar \chi\,,
\ee
and  where $\delta k^\mu$ is computed using the geodesic equation for the comoving null geodesic vector $k^\mu (\chi)=(\ud x^\mu/\ud \chi)(\chi) $.
More precisely,
\begin{eqnarray}
\frac{\ud k^\mu (\chi) }{ \ud \chi}+\hat \Gamma^\mu_{\alpha \beta}(x^\gamma)\, k^\alpha(\chi) \,  k^\beta(\chi) =0\,,
\end{eqnarray}
where $\hat \Gamma^\mu_{\alpha \beta}$ are the Christoffel symbols defined using the comoving metric $\hat g_{\mu  \nu}= g_{\mu  \nu}/a^2$.
Expanding $k^\mu (\chi)$ and $\hat \Gamma^\mu_{\alpha \beta}(x^\gamma)$ up to linear order:
\begin{eqnarray}
\label{photon-gamma-pert}
 k^\mu (\chi) &=& k^\mu (\bar\chi)  + \delta\chi  \frac{\ud k^\mu }{ \ud \chi}(\bar \chi) \;, \nonumber \\
 \hat \Gamma^\mu_{\alpha \beta}( x^\gamma) &=&\hat \Gamma^\mu_{\alpha \beta}(\bar x^\gamma) + \Delta x^\nu \frac{\p }{\p \bar x^\nu} \hat \Gamma^\mu_{\alpha \beta}(\bar x^\gamma)\;,
\end{eqnarray}
we get
\begin{eqnarray}
\label{Eqgraviton}
\frac{\ud k^\mu (\bar\chi) }{ \ud \bar \chi}+ \hat \Gamma^\mu_{\alpha \beta}(\bar x^\gamma) k^\alpha(\bar\chi)  k^\beta(\bar\chi) =0 \;,
\end{eqnarray}
where $k^\mu(\bar\chi) = \bar k^\mu (\bar\chi)  + \delta  k^\mu (\bar\chi) $.\\
We then need to evaluate the scale factor and affine parameter at emission.
For the former,
\be\label{a}
\frac{a}{\bar a} = 1 + \Delta \ln a=1+\cH\Delta x^{0 },
\ee
where $\bar a = a (\bar x^0)$  and $\cH=\bar a'/\bar a$  is the conformal Hubble factor. Here the prime is $\p/\p \bar x^0 = \p/\p \bar \eta$. As show in Section \ref{sec:firstorder}, $\bar a = 1/(1+z)$. Now, we have
\be
\label{affine}
\ud \chi =\left( 1+{\ud \delta \chi \over \ud \bar \chi}\right)  \ud  \bar \chi \;,
\ee
with\footnote{Here we used that $\delta \chi  =    \delta x^{0}- \Delta x^{0 } = \delta x^{0} - \Delta \ln a/\cH$.}
\begin{equation}
\label{dchi_1/chi}
{\ud \delta \chi \over \ud \bar \chi} =\delta  f -{\cH'\over \cH^2}\Delta \ln a - \frac{1}{\cH}{\ud \Delta \ln a  \over \ud \bar \chi} \;.
\end{equation}

It then remains to study the total density, which depends on halo mass, the environment around the halo, e.g. tidal effects, velocity dispersion, type of galaxies. Most of these effects could change not only the background number
density of the halos but also the relation between the density contrast
of the halos and dark matter.
It is therefore essential to have {\it a priori } the knowledge of an astrophysical model that connects all these quantities, e.g. see \cite{Flanagan:1997sx, Ferrari:1998jf, Schneider:1998xt,  Schneider:2000sg,   TheLIGOScientific:2016wyq} (see also \cite{Regimbau:2011rp} and  refs. therein).
Perturbing the total density, we get\footnote{Note that
$$n^{[i]}(x^\alpha)= n^{[i](0)}(x^0) +  n^{[i](1)}(x^\alpha), \quad {\rm and} \quad n^{[i](0)}(x^0)  = n^{[i](0)}(\bar x^0 + \Delta x^0) =  \bar n^{[i]} (\bar x^0) +  \frac{\p \bar n^{[i]} }{\p \bar x^0} \, \Delta x^{0}\;,$$
where $n^{[i](0)}(\bar x^0)=\bar n^{[i]} (\bar x^0)$.}
\begin{equation}
n^{[i]}(x^\alpha) =  \bar n^{[i]} \Bigg(1 + \frac{\ud \ln \bar n^{[i]} }{\ud  \ln \bar a} \, \Delta \ln a + \delta^{[i]}   \Bigg) \;,
\end{equation}
where
\begin{equation}
\delta^{[i]}= {n^{[i](1)}(\bar x^\alpha) \over \bar n^{[i]} (\bar x^0)} \;.
\end{equation}
We thus obtain
\be
\Omega_{\rm GW}= \frac{ f_\ro}{\rho_{\rm c}} \frac{\ud \rho_{\mathrm{GW}}}{\ud  f_\ro \ud \Omega_\ro} ={ \bar \Omega_{\rm GW}\over 4\pi}+ \Delta  \Omega_{\rm GW},
\ee
where
\be \label{back-rho_GW}
{ \bar \Omega_{\rm GW}\over 4\pi} =\frac{ f_\ro}{\rho_{\rm c}} \frac{\ud \bar \rho_{\rm GW}}{\ud  f_\ro \ud \Omega_\ro} = \frac{ f_\ro}{\rho_{\rm c}}  \sum_{[i]} \int  {N^{[i]}(z,  f_\re, \vec{\theta} )\over  (1+z)}  \; \ud  \bar \chi \,  \ud \vec{\theta} \;,
\ee
with $ N^{[i]}(z, f_\re, \vec{\theta})  = \bar n^{[i]}(z, f_\re, \vec{\theta}) /  (1+z)^3 $ the {\it total} comoving number density at a given redshift,
and
\begin{eqnarray}\label{DeltaOmega}
 \Delta \Omega_{\rm GW} = \frac{ f_o}{\rho_c}  \sum_{[i]} \int   {N^{[i]}(z, f_\re, \vec{\theta}) \over  (1+z)} \Bigg\{\delta^{[i]} +  \frac{\ud \ln  N^{[i]}  }{\ud  \ln \bar a} \, \Delta \ln a - \left(1 + {\cH'\over \cH^2} \right) \Delta \ln a +\delta  f  - \frac{1}{\cH}{\ud \Delta \ln a  \over \ud \bar \chi}\Bigg\}  \; \ud  \bar \chi  \,  \ud \vec{\theta}  \;.
 \end{eqnarray}

\subsection*{Connection with  Halo and Stellar mass functions and with Star Formation Rate}

 It is important to relate the above quantities  to the halo and stellar mass function, and to the Star Formation Rate (SFR).
 At background level, for each type of source,  in literature  the comoving rate density is defined as \cite{Phinney:2001di}

 \be
 R^{[i]} (z, \vec{\theta}) \equiv {1\over (1+z)} \frac{\ud N_{\rm GW}^{[i]}(z, \vec{\theta}) }{\ud \T_\re} \chi^2 {\ud \chi \over \ud z}\ud \Omega_\re \,,
 \ee
 where $N_{\rm GW}^{[i]}$ is the comoving number density of ASGW.
Precisely, $ N_{\rm GW}^{[i]}(z, \vec{\theta})$ depends on the  mass of  stars\footnote{In principle, $N_{\rm GW}^{[i]}$ should be function on the stellar mass at given $z$ and $M^*$, e.g. see \cite{Nakazato:2016nkj, Maiolino:2008gh}.  Finally, more in general, we could split this quantity in three parts: i) contribution of central galaxy, (ii) satellite galaxies and (iii) all sources that are still within the halo, but outside the host galaxies.} $ M^*$ that give origin to the sources that we are considering, i.e.
 \be
 \ud N_{\rm GW}^{[i]}= \frac{\ud  N_{\rm GW}^{[i]} }{\ud  M^*} \ud M^*\;.
 \ee
  Following  \cite{Behroozi:2010rx,Behroozi:2012iw}, the stellar mass $M^*(M_{\rm h})$  is a function of host halo mass $M_{\rm h}$ [in general, it could also depend on many other parameters as the metallicity (e.g. see \cite{Nakazato:2016nkj}), etc.]. Then we have \cite{Behroozi:2010rx}
 \be
 \frac{\ud N_{\rm GW}^{[i]}}{\ud \ln M^*} =   \frac{\p N_{\rm GW}^{[i]}}{\p \ln M_{\rm h}}  \left({\ud \log_{10} M^* \over \ud \log_{10} M_{\rm h}} \right)^{-1}  \;.
 \ee
 For a given halo mass $M_{\rm h}$ we can split  $N^{[i]}_{\rm GW}(\vec \theta^*, M_{\rm h},M^*,\vec m,\T_\re,z)=N_{\rm h}(M_{\rm h},z) \langle{{\mathcal N}^{[i]}_{\rm GW}}(\vec \theta^*, M^*, \vec m, z, \T_\re)\rangle$,
 where $N_{\rm h}$  is the comoving number density of halos in a mass interval  $\ud M_{\rm h}$ around $M_{\rm h}$.
   Comparing these relations with the background quantities (in the observed frame), described in the previous sections, we find
  \be
  N^{[i]}(z, \vec{\theta}) =\lambda^{[i]}(z, \vec{\theta}) \frac{ \ud N_{\rm h} (M,z) }{\ud  M_{\rm h}} \quad {\rm and}  \quad   {{\mathcal N}^{[i]}_{\rm GW}}_\re  = \langle{{\mathcal N}^{[i]}_{\rm GW}}(\vec \theta^*, M^*, \vec m, z, \T_\re)\rangle  \;,
  \ee
  where $\lambda^{[i]}(z, \vec{\theta})$ is a generic function which depends on the initial mass function $M^*$ and, in general, other parameters of the sources. Because of the many simplifications we have taken up here,  we define
  \be
 \lambda^{[i]}(z, \vec{\theta})= \frac{M_{\rm h}}{M^*} \left({\ud \log_{10} M^* \over \ud \log_{10} M_{\rm h}} \right)^{-1} \bar {\mathcal K}^{[i]}(z, f_e,  \vec{\theta} ) \;.
  \ee
 Then we have
 \be
 \frac{\p N_{\rm GW}^{[i]}(z, \vec{\theta})}{\ud \T_\re \p  M_{\rm h}}=  \frac{\ud N_{\rm h} (M_{\rm h},z) }{\ud  M_{\rm h}} \frac{\ud  {{\mathcal N}^{[i]}_{\rm GW}}_\re }{\ud \T_\re}\;.
 \ee
 Now $N_{\rm h}(M_{\rm h},z)$ can be related to the fraction of mass $F(M_{\rm h},z)$ that is bound at the epoch $z$ in halos of mass smaller than $M_{\rm h}$, i.e.
 \be
 \frac{\ud N_{\rm h} (M_{\rm h},z) }{\ud  M_{\rm h}} = {\bar \rho(z)\over M_{\rm h}}  \frac{\ud F (M_{\rm h}) }{\ud  M_{\rm h}}\;,
 \ee
 where $\bar \rho(z)$ is the comoving background density.  How, for example, we can use the Press \& Schechter (1974) \cite{Press:1973iz} the Sheth \& Tormen  (1999) \cite{Sheth:1999mn} or the Tinker (2008) \cite{Tinker:2008ff} mass fraction. Following \cite{Springel:2002ux, Hernquist:2002rg}, it is useful to define $g(M)$ of halos
 \be
g(M_{\rm h})= \frac{\ud F (M_{\rm h},z) }{\ud \ln M_{\rm h}}\;.
 \ee
Finally let us introduce the (mean) star formation rate  SFR that it is connected with ${{\mathcal N}^{[i]}_{\rm GW}}_e$ in the following way
\be
\frac{\ud  {{\mathcal N}^{[i]}_{\rm GW}}_\re }{\ud \T_\re}=   {{\mathcal N}^{[i]}_{\rm GW}}_\re \times {\rm SFR}\;.
\ee
Note that $s(M_{\rm h},z)$ defined in \cite{Springel:2002ux, Hernquist:2002rg} can be related with ${\rm SFR} $ in the following way  $s(M_{\rm h},z)= (M^*/M_h)\times{\rm SFR}$.
We conclude that this analysis can be easily used for case (I). Nonetheless we can use the above approach also for the case (II). Indeed, if we define the following new quantity
\be
\mathcal F = \frac{\ud \ln  \left({\mathcal{E}^{[i]}_{\rm GW}}_\re / \ud  f_\re   \ud A_\re \right)} {\ud \T_\re}
\ee
then, substituting SFR with $\mathcal F$, we can use again the above prescription.

\section{First order metric terms}
\label{sec:firstorder}

Let us now consider a spatially flat FLRW background, perturbed in a general gauge at the linear order:
 \begin{eqnarray}
 \label{metric}
 \ud s^2 = a(\eta)^2\left[-\left(1 + 2A\right)\ud\eta^2-2 B_{i}\ud\eta\ud x^i+\left(\delta_{ij} +h_{ij}\right)\ud x^i\ud x^j\right] \;,
\end{eqnarray}
where $B^{i}= \p_i B + \hat B_i$, with $ \hat B_i$ a solenoidal vector, i.e. $\p^i\hat B_i=0$, and
$h_{ij} = 2 D\delta_{ij} + F_{ij}$, with $F_{ij}= (\p_i\p_j- \delta_{ij}\nabla^2/3)F+\p_i \hat F_j+ \p_j \hat F_i+\hat h_{ij}$. Here $D$ and $F$ are scalars and $\hat F_i$ is a solenoidal vector field,   $\p^i \hat h_{ij}=\hat h_i^{i}=0$.

Considering a  four-velocity vector $u^\mu$
at linear order,
\begin{eqnarray}
\label{u0i}
u_0=-a\left(1+A\right) , \quad
u_i=a\left(v_i -B_i \right), \quad
\end{eqnarray}
and using Eqs.\ (\ref{E0mu}), (\ref{LambdaE}) and \eqref{u0i}, we can deduce all components of $\Lambda_{\hat a \mu}^{(n)}$ and $E_{\hat a \mu}^{(n)}$, as follows:
\be \label{LambdaE-0}
E_{\hat{0}\mu}^{(0)}=(-1, {\bf 0})\;, \quad \quad{ \rm and} \quad\quad E_{\hat{0}}^{\mu(0)}=(1, {\bf 0})
\ee
 at background level and
\begin{eqnarray} \label{LambdaE-1}
\begin{array} {lll}
\Lambda_{\hat 0 }^{0(1)}= E_{\hat 0 }^{0(1)}/a= -A/a \;, & \quad& \Lambda_{\hat 0 }^{i(1)}= E_{\hat 0 }^{i(1)}/a= v^i/a\;,  \\  \\
\Lambda_{\hat 0 0}^{(1)}=a E_{\hat 0 0}^{(1)}= -a A \;, & \quad& \Lambda_{\hat 0 i}^{(1)}=a E_{\hat 0 i}^{(1)}= a \left(v_i -B_i \right)\;,  \\  \\
\Lambda_{\hat a 0}^{(1)}=a E_{\hat a 0}^{(1)}= -a v_{\hat a} \;, & \quad& \Lambda_{\hat a i}^{(1)}=a E_{\hat a i}^{(1)}=\frac{1}{2} a h_{\hat a i}\;.  \\  \\
\end{array} 
\end{eqnarray}
at first order. The geodesic equation (\ref{Eqgraviton}) yields\footnote{This is in agreement with photon perturbation analysis made (see for example \cite{Jeong:2011as, Schmidt:2012ne}).}
\begin{eqnarray}
\label{dnu-1}
\frac{\ud}{\ud\bar \chi} \left(\delta  f - 2 A +  B_\| \right) &=& A{'} - B_\|{'} - \frac{1}{2} h_{\|}^{'} \\
\label{de-1}
\frac{\ud}{\ud\bar \chi} \left( \delta n^{i} + B^{i} + h_j^{i} n^j \right) &=&
- \p^i A + \p^i B_\|- \frac{1}{\bar \chi}B_{\perp}^{i} + \frac{1}{2} \p^i h_\| - \frac{1}{\bar \chi} \Perp^{ij} h_{jk} n^k \;,
\end{eqnarray}
From Eq.\ (\ref{p_GW}), for $\bar \chi=0$ we have
\be
p^{\mathrm{GW}}_{\hat{0} \ro} =( \Lambda_{\hat{0} \mu}p_{\mathrm{GW}}^\mu)|_\ro=-2\pi f_\ro\;, \quad \quad
p^{\mathrm{GW}}_{\hat{a}  \ro} = (\Lambda_{\hat{a} \mu}p_{\mathrm{GW}}^\mu)|_\ro= -2\pi  f_\ro n_{\hat{a}}\;,
\ee
and we find
\bea
( \Lambda_{\hat{0} \mu}p_{\mathrm{GW}}^\mu)|_\ro= - {2\pi  f_\ro \over a_\ro} (E_{\hat{0} \mu} k^\mu)|_\ro=-2\pi f_\ro\;, \label{nu_o}\;\;\;\;\;\;\;\;\;(\Lambda_{\hat{a} \mu}p_{\mathrm{GW}}^\mu)|_\ro = - {2\pi  f_\ro \over a_\ro}  (E_{\hat{a} \mu}k^\mu)|_\ro= -2\pi  f_\ro n_{\hat{a}}\,.
\eea
Using Eqs. (\ref{kmu}), (\ref{LambdaE-0}), (\ref{LambdaE-1}) and
\be \label{a_o}
a_\ro = a(\eta_0)=  \bar a (\bar \eta_0) + \delta a_\ro = 1 + \delta a_\ro \;
\ee
where we set $\bar a (\bar \eta_0)=1 $, at the observer we have
\begin{eqnarray}
\label{dnude-1o}
\delta f_\ro&=&-\delta a_\ro +A_\ro+v_{\| \ro}-B_{\| \ro}\;, \\
\delta n^{\hat a }_\ro&=&\delta a_\ro n^{\hat a}  -v^{\hat a }_\ro-\frac{1}{2} n^i h^{\hat a }_{i \, \ro} \;.
\end{eqnarray}
From Eqs.\ (\ref{dnu-1}), (\ref{de-1}) and the constraint from Eq.\ (\ref{dnude-1o}), we obtain at first order with
\begin{eqnarray}
\label{dnude-1}
\delta f
&=& -\delta a_\ro  - \left (A_\ro-v_{\| \, \ro}\right)+ 2 A - B_{\| } - 2I \\
\delta n^{i}
&=& n^i \delta n_\|^{ (1)}+\delta n_\perp^{i}\;,
\end{eqnarray}
where
\begin{eqnarray}
\label{dnue-||perp-1}
\delta n_\| &=&\delta a_\ro + A_\ro-v_{\| \, \ro}-A- \frac{1}{2}  h_{\| }+2I  \quad \quad \quad \\
\delta n_\perp^{i}&=&  B^{i}_{\perp \, \ro }-v^{i}_{\perp \, \ro }+ \frac{1}{2} n^k h_{k\,\ro}^{j(1)} \Perp^i_j-\left( B^{i}_{\perp }+ n^k h_{k}^{j(1)} \Perp^i_j\right) + 2S_{\perp}^{i(1)}  \;,
\end{eqnarray}
and
\begin{eqnarray}
\label{iota}
I& \equiv& -\frac{1}{2} \int_0^{\bar \chi} \ud \tilde \chi \left(A{'} - B_{\| }{'} - \frac{1}{2}h_{\| }{'} \right) \quad \quad \quad \\
S^{i} &\equiv& -\frac{1}{2} \int_0^{\bar \chi} \ud \tilde \chi \left[ \tilde\p^i \left( A - B_{\| } - \frac{1}{2}h_{\| } \right) + \frac{1}{\tilde \chi} \left(B^{i}+  n^k h_{k}^{i} \right)\right]\;;
\label{varsigma}
\end{eqnarray} $I$ is the Integrated Sachs-Wolfe contribution and $\tilde \p_i\equiv \p /\p \tilde x^i$.

We can use the projector parallel and perpendicular to the line of sight, defined in Appendix \ref{A}, to split  $S^{i} $ in its parallel and perpendicular components
\begin{eqnarray}
S_{\perp}^{i} &=& -\frac{1}{2} \int_0^{\bar \chi} \ud \tilde \chi \left[ \tilde\p^i_\perp \left( A - B_{\| } - \frac{1}{2}h_{\| } \right) + \frac{1}{\tilde \chi} \left(B^{i (n)}_{\perp }+  n^k h_{kj} \Perp^{ij}  \right)\right]\;,\\
S_{\|} &=& \frac{1}{2} \left( A_\ro - B_{\| \, \ro} - \frac{1}{2}h_{\| \, \ro} \right)- \frac{1}{2} \left( A - B_{\| } - \frac{1}{2}h_{\| } \right)+ I- \frac{1}{2} \int_0^{\bar \chi} \ud \tilde \chi \frac{1}{\tilde \chi} \left( B_{\| }+h_{\| } \right)\;.
\end{eqnarray}
Note the relation
\begin{eqnarray}
\label{du+de}
\delta n_\| +  \delta f&=&A-B_{\| }- \frac{1}{2} h_{\| }\;.
\end{eqnarray}

Using Eqs. (\ref{delta-x}), (\ref{dnude-1}) and  (\ref{dnue-||perp-1}), we find at first order
\begin{eqnarray}
\label{dx0-1}
\delta x^{0}&=&\delta x^{0}_\ro  -\bar \chi \left(\delta a_\ro+ A_\ro-v_{\| \, \ro}\right)+ \int_0^{\bar \chi} \ud \tilde \chi \left[ 2 A - B_{\| } + \left(\bar \chi-\tilde \chi\right) \left(A{'} - B_{\| }{'} - \frac{1}{2}h_{\| }{'} \right) \right] \; \\
\label{dx||-1}
\delta x_{\|}&=& \delta x_{\| \ro}+\bar \chi \left(\delta a_\ro+ A_\ro-v_{\| \, \ro}\right) -\int_0^{\bar \chi} \ud \tilde \chi \left[ \left(A+ \frac{1}{2}  h_{\| }\right) +  \left(\bar \chi-\tilde \chi\right) \left(A{'} - B_{\| }{'} - \frac{1}{2}h_{\| }{'} \right) \right] \;,   \\
\label{dxperp-1}
\delta x_{\perp}^{i (1)}&=&\delta x_{\perp \ro}^{i}+ \bar \chi \left(B^{i}_{\perp \, \ro }-v^{i}_{\perp \, \ro }+ \frac{1}{2} n^k h_{k\,\ro}^{j} \Perp^i_j\right)-\int_0^{\bar \chi} \ud \tilde \chi \left\{ \left( B^{i }_{\perp }+ n^k h_{k}^{j} \Perp^i_j\right) \right. \nonumber \\
&&+\left .\left(\bar \chi-\tilde \chi\right)\left[\tilde \p^i_\perp \left( A - B_{\| } - \frac{1}{2}h_{\| } \right) + \frac{1}{\tilde \chi} \left(B^{i}_{\perp }+  n^k h_{kj} \Perp^{ij}  \right)\right] \right\}\;,\\
\label{s-1}
\delta x^{0(1)} + \delta x_{\|}^{(1)} &=& \delta x^{0(1)}_\ro + \delta x_{\| \ro}^{(1)}  -  T \;,
\end{eqnarray}
where
\be
T = -   \int_0^{\bar \chi} \ud \tilde \chi \left(A - B_{\| }- \frac{1}{2}h_{\| }\right)\,,
\ee
is the Shapiro time delay \cite{PhysRevLett.13.789}.\\

The quantities $\delta x^{0}_\ro$ and $\delta x^{i}_\ro$, derived in the Appendix \ref{B} following \cite{Fanizza:2018qux, Biern:2016kys}, have their origin from the fact that the physical coordinate time $t_0=t(\eta=\eta_0)=t_{\rm in}+\int_{\eta_{\rm in}}^{\eta_0} a(\tilde \eta) \ud\tilde \eta$ does not coincide with the proper time of the observer $\T_0$ in an inhomogeneous universe. We have
\be
\delta x^{0}_\ro=\delta \eta_0= \int_{\bar \eta_{\rm in}}^{\bar \eta_0} \bar a \; E_{\hat 0}^0 \ud \tilde \eta=- \int_{\bar \eta_{\rm in}}^{\bar \eta_0} \bar a(\tilde \eta) \; A(\tilde \eta, {\bf 0})\; \ud \tilde \eta
\ee
and
\bea
\delta x^{i}_\ro &=&  \int_{\T_{\rm in}}^{\T_0}  \delta u^i \ud \tilde \T=  \int_{\bar \eta_{\rm in}}^{\bar \eta_0} v^i (\tilde \eta, {\bf 0})\; \ud \tilde \eta\;.
\eea
Taking into account that $a(\bar \eta_0 +\delta \eta_\ro )=1+H_0 \delta \eta_\ro =1+\delta a_\ro$, we are able to obtain the expression for $\delta a_0$$
=1+\delta a_0$,
\be
\delta a_\ro = - H_0 \int_{\bar \eta_{\rm in}}^{\bar \eta_0} \bar a(\tilde \eta) A(\tilde \eta, {\bf 0})\; \ud \tilde \eta\;.
\ee

The next quantity that we need  to compute explicitly is $\Delta \ln a$. The observed redshift is given by
\begin{eqnarray} \label{z}
(1+z) = \frac{(u_\mu p_{\rm GW}^\mu)\big|_{\re} }{(u_\mu p_{\rm GW}^\mu)|_\ro}=\frac{a_\ro}{a(\chi_\re)}\frac{(E_{\hat{0}\mu} k^\mu)\big|_\re }{(E_{\hat{0}\mu} k^\mu)|_\ro}\;,
\end{eqnarray}
where we used $ f \propto 1/a$. Quantities evaluated at the observer have a subscript o, while other quantities are assumed to be evaluated at the emitter (up to first order).
As we discusses above in Eqs.\ (\ref{nu_o}), (\ref{a_o}) and (\ref{dnude-1o}), we know that
\begin{eqnarray}
\label{Ek_o}
(E_{\hat{0}\mu} k^\mu)|_o=1-\delta  f_o +A_o +v_{\| o} - B_{\|o} =1+\delta a_o =a_o \;;
\end{eqnarray}
then we have
\begin{eqnarray}\label{z2}
1+z=\frac{E_{\hat{0}\mu} k^\mu}{a}\;.
\end{eqnarray}
From Eq.\ (\ref{a}), $\bar a = 1/(1+z)$ is the scale factor of the observed frame. From Eqs.\ (\ref{a}),  (\ref{LambdaE-0}) and (\ref{LambdaE-1}), ({\ref{z2}}) turns out to be
\begin{eqnarray}
\label{Ek-total}
1=\frac{1+ (E_{\hat{0}\mu} k^\mu)}{ 1 + \Delta \ln a}\;,
\end{eqnarray}
where
\begin{eqnarray}
(E_{\hat{0}\mu} k^\mu)^{(0)} = 1\;.
\end{eqnarray}
Then from Eq.\ (\ref{Ek-total}) we can find $\Delta \ln a $, such that
\bea
\label{Ek-1}
\Delta \ln a &=& (E_{\hat{0}\mu} k^\mu)^{(1)} = E_{\hat{0}\mu}^{(1)} k^{\mu (0)}+E_{\hat{0}\mu}^{(0)} k^{\mu (1)}=- E_{\hat{0}0}^{(1)} + n^i E_{\hat{0}i}^{(1)} - \delta  f=A+v_{\| } - B_{\|}- \delta  f
\nonumber\\
&=&\delta a_\ro+ \left (A_\ro-v_{\| \, \ro}\right) - A+ v_\|+ 2I\;.
\eea
Note that this result was already obtained for the photon in \cite{Jeong:2011as, Schmidt:2012ne}. In this case we are able to write explicitly Eqs. (\ref{chi}) and (\ref{dchi_1/chi}), and obtain the final equation for the affine parameter Eq.~(\ref{affine})

\bea
\delta \chi
&=&
\delta x^{0}_\ro  - \left( \bar \chi+\frac{1}{\cH}\right)\left(\delta a_\ro+A_\ro-v_{\| \, \ro}\right)+  \frac{1}{\cH}\left(A- v_\| \right) \nonumber \\
&& + \int_0^{\bar \chi} \ud \tilde \chi \left[ 2 A - B_{\| }+\left(\bar \chi-\tilde \chi\right) \left(A{'} - B_{\| }{'} - \frac{1}{2}h_{\| }{'} \right) \right] -\frac{2}{\cH}I \;
\eea
and
\begin{eqnarray}
\label{ddeltachi/dchi}
{\ud \delta \chi \over \ud \bar \chi}  &=&- \left( 1+\frac{\cH'}{\cH^2}\right)\left(\delta a_\ro+A_\ro-v_{\| \, \ro}\right) + \left( 2 +\frac{\cH'}{\cH^2}\right) A -   B_{\| }-\frac{\cH'}{\cH^2} v_\|  +\frac{1}{\cH}\left[\frac{\ud \,}{\ud \bar \chi} \left( A -  v_\| \right)+ \left(A{'} - B_{\| }{'} - \frac{1}{2}h_{\| }{'} \right)\right] \nonumber\\
&&- 2 \left( 1+\frac{\cH'}{\cH^2}\right) I \;.
\end{eqnarray}

Hence it reads
\begin{eqnarray}
\label{DeltaOmega2}
 \Delta \Omega_{\rm GW} & =&  \frac{ f_\ro}{\rho_{\rm c}}  \sum_{[i]} \int{N^{[i]}[z, f_\ro (1+z)]\over  (1+z)}    \; \Bigg\{ \delta^{[i]} + \left(- b^{[i]}_\re  + 3 +  {\cH'\over \cH^2}\right) A-   B_{\| }   \nonumber\\
 &&+ \left(b^{[i]}_\re -1 - {\cH'\over \cH^2}\right)  v_\|   +\frac{1}{\cH}\left[\frac{\ud \,}{\ud \bar \chi} \left( A -  v_\| \right)+ \left(A{'} - B_{\| }{'} - \frac{1}{2}h_{\| }{'} \right)\right] + 2\left(b^{[i]}_\re - 2 - {\cH'\over \cH^2}\right)  I  \nonumber\\
 &&+ \left(b^{[i]}_\re -2 - {\cH'\over \cH^2}\right) \left(\delta a_\ro+ A_\ro-v_{\| \, o}\right)\Bigg\}  \; \ud  \bar \chi \;,
 \end{eqnarray}
 where we have defined the evolution bias by
\begin{eqnarray}
b^{[i]}_\re = \frac{\ud \ln N^{[i]}}{\ud \ln \bar a}=-\frac{\ud \ln N^{[i]}}{\ud \ln (1+z)}.
\end{eqnarray}
Note that, in general, $\delta^{[i]} $ is not a gauge invariant quantity.


\section{Gravitational wave background anisotropy in the Synchronous-Comoving gauge} \label{sec:aps}

Using the synchronous-comoving (SC) gauge within $\Lambda$CDM allows us to synchronise observers on the same spacelike hypersurface, as they are comoving with the cosmic expansion. The metric can be then written as
\begin{equation}\label{sync}
ds^2=a^2(\eta)\Big\{-d\eta^2+\Big[\big(1-2{\mathcal R}
\big)\delta_{ij}+2\partial_i \partial_j E\Big] dx^idx^j\Big\}\,,
\end{equation}
where  as previously $\eta$ denotes conformal time and we set   $g_{00}=-1$,   $g_{0i}=0$ and $v^{i}=0$. Hence, $A=0$, $B_i=0$, $F=2E$ and ${\mathcal R} = -D+\nabla^{2}E/3$ (or $h_{ij}=-2{\mathcal R}\delta_{ij}+2 \partial_i \partial_j E $).

In the SC gauge the bias $\delta^{[i] \rm (SC)} $ is a gauge invariant quantity.
Moreover, in SC gauge, the spherical collapse model has an exact GR interpretation and only in this frame halos collapse when the linearly growing local density contrast (smoothed on the corresponding physical mass scale) reaches a critical value.
Quantitatively,
which on large scales can be defined as
\begin{equation}\label{def-bias}
\delta^{[i] \rm (SC)} = b^{[i]} (\eta) \delta_{\rm m}^{\rm (SC)}\;.
\end{equation}
To simplify notation, in what follows we drop the superscript ``SC", but still use the SC gauge unless explicitly specified otherwise.

We thus obtain
\begin{eqnarray}
\label{DeltaOmegaSC}
 \Delta \Omega_{\rm GW}({\bf n}) &=& \frac{ f_\ro}{\rho_{\rm c}}  \sum_{[i]} \int {N^{[i]} [z, f_\ro (1+z)]\over (1+z)}  \times \nonumber \\
 &\times& \Bigg\{  b^{[i]} (\eta) \delta_{\rm m}  -\frac{1}{\cH} \p_\|^2 E' + \left(b^{[i]}_\re - 2 - {\cH'\over \cH^2}\right) \left( \partial_\| E'+ E'' \right)\Big|_\ro^{\bar \chi}+ \left(b^{[i]}_\re - 2 - {\cH'\over \cH^2}\right) \int_0^{\bar \chi}  E''' \ud \tilde \chi  \Bigg\}  \; \ud  \bar \chi \;.
 \end{eqnarray}
The physics behind the different contributions is  clear: there are local terms taking into account the evolution from source to the observer,  including the galaxy density perturbation (the first term within the curly brackets), the Kaiser term (i.e. $ -(1/\cH)\p_\|^2 E'$), the Doppler effect  (i.e. proportional $ \partial_\| E'$ term), the local gravitational potential term  (proportional to  $ E''$), and finally the Integrated Sachs-Wolfe contribution (proportional to  $ \int_0^{\bar \chi}  E''' \ud \tilde \chi$). 

While the structure is similar to the one found in~\cite{Cusin:2017fwz, Cusin:2017mjm, Jenkins:2018lvb, Cusin:2018rsq, Jenkins:2018uac}, our result is expressed in the observer's frame, which, by definition makes all quantities gauge invariant.  We can then evaluate the evolution bias related to the distribution of objects along the line-of-sight.
Note that since we are working in the observer's frame, we do not need to perturb the effective luminosity.
Nevertheless, for completeness we also present our result in the Poisson gauge In Appendix \ref{B}.

\section{ Angular power spectrum} \label{sec:AngulaPS}
To characterize the ASGWB we compute the correlation between the energy density coming from different directions. It is known that this is the appropriate quantity to correlate~\cite{Allen:1997ad,Romano:2016dpx}, rather than the GW signal itself, which would have a vanishing two point correlation, unless signals with coherent phases are considered. Since we measure it on a two dimensional sky, the spherical symmetry allows us to work in spherical harmonics space. Therefore, we expand the observed GW energy density as
\begin{align}
\Delta \Omega_{\rm GW}({\bf n})  &=\sum_{\ell m} {\alpha}_{\ell m} Y_{\ell m}({\bf n}) \;,
\end{align}
where the coefficients ${\alpha}_{\ell m}$ are given by
\be
\alpha_{\ell m} = \int\!\!\ud^2{\bf n}\, Y^{*}_{\ell m}({\bf n}) \Delta \Omega_{\rm GW}({\bf n})\;.
\ee
The angular power spectrum then reads
\begin{align}
\label{eq:clgg}
C^{\rm GW}_{\ell}&\equiv \sum_{\ell=- m}^{\ell=m}\frac{\left\langle {\alpha}^*_{\ell m}~
{\alpha}_{\ell m} \right\rangle}{2\ell+1}=\sum_{i,j;\alpha,\beta} C^{[ij]\alpha\beta}_\ell\,
\end{align}
where
 \be
 C^{[ij]\alpha \beta}_\ell = \sum_{\ell=- m}^{\ell=m}\frac{\left\langle {\alpha}^{[i]\alpha*}_{\ell m}~
{\alpha}^{[j]\beta}_{\ell m} \right\rangle}{2\ell+1}=\int {k^2\over(2\pi)^3} {\mathcal S^{[i]\alpha}_{\ell}}^*(k) \mathcal S^{[j]\beta}_{ \ell}(k) P_{\rm m}(k) \, \ud k \, ,
\ee
with $P_m$ the matter power spectrum today
\begin{equation}
\langle \delta({\bf k}, \eta_0) \delta^*({\bf k'}, \eta_0) \rangle
= (2\pi)^3 \delta^{(3)}_D ( {\bf k}-{\bf k}' ) P_m(k)\;.
\end{equation}
Therefore in spherical space
\be
{\alpha^{[i]\alpha}}_{\ell m} =   \int  \frac{{\rm d}^3  { \bf k}}{(2\pi)^3}\, Y^*_{\ell m}(\hat{\bf k})   {\mathcal S}^{[i] \alpha}_{\ell}(k)  \delta_{\rm m}({ \bf k},\eta_0)\;,
\ee
where we defined the spherical transforms as
\begin{eqnarray}
{\mathcal S}_{\ell}^{[i]\alpha}(k) \equiv 4 \pi i^{\ell} \int {\rm d}{\bar \chi}   ~ { \mathcal W}^{[i]}( \bar\chi)\int_0^{\bar \chi} {\rm d} \tilde \chi ~\left[\mathbb{ W}^{\alpha} \left(\bar \chi, \tilde \chi, \eta, \tilde \eta, \frac{\partial}{\partial \tilde \chi}, \frac{\partial}{\partial \tilde \eta}\right)  {\Upsilon}^{\alpha}( {\bf k},\tilde \eta) j_{\ell} ( k \tilde \chi)\right] \; ,
\end{eqnarray}
with
\be
{ \mathcal W}^{[i]}( \bar\chi(z)) =\frac{ f_o}{\rho_c}   {N^{[i]}[z, f_\ro (1+z)] \over  (1+z)} \;.
\ee
For each contribution in Eq.~\eqref{DeltaOmegaSC} we define the operator $ \WW^{\alpha}$, which encloses the different physical effects, and
${\Upsilon}^{\alpha}( {\bf k},\tilde \eta)$ is a transfer function that maps the different perturbed contributions at a given redshift to the density contrast today.
Precisely, in $\Lambda$CDM, taking into account that $E''+aHE'-4\pi G a^2 \rho_m  E =0$ (note that ${\mathcal R}'=0$), we have
\begin{eqnarray}
E'     &=&-\frac{H}{(1+z)}f\nabla^{-2}\delta_{\rm m}, \\
E''    &=&-\frac{H^2}{(1+z)^2} \Big(\frac{3}{2} \Omega_{\rm m}-f\Big)\nabla^{-2}\delta_{\rm m}, \\
E'''   &=&-3 \frac{H^3}{(1+z)^3}\Omega_{\rm m} \left(f-1\right)\nabla^{-2}\delta_{\rm m}\, ,\\
{\mathcal R} &=&  \frac{H^2}{(1+z)^2} \Big(\frac{3}{2} \Omega_{\rm m}+f\Big) \nabla^{-2}\delta_{\rm m}\,.
\end{eqnarray}
Here $\Omega_m(z)$ is the matter density and  $f(z)$ is the growth rate defined as
\begin{equation}
f={d\ln D \over d\ln a}\,, \quad \quad \delta_{\rm m}({\bf x},\eta)= \delta_{\rm m}^{\rm (SC)}({\bf x},\eta_0){D(\eta) \over D(\eta_0)}\,,
\end{equation}
where $D $ is the growing mode of $\delta_{\rm m}^{\rm (SC)}$. In conclusion
The ${\mathcal S}_{\ell}^a(k)$ functions describe the different physical effects and can be written as:
\begin{eqnarray}
&&{\mathcal S}_{\ell }^{[i] \delta_{\rm m}^{\rm (SC)} }(k)= (4 \pi) i^{\ell} \int {\rm d} \bar \chi ~ { \mathcal W}^{[i]}( \bar\chi)  b_{\mathrm{gw}}^{[i]} (\eta) {D(\eta) \over D(\eta_0)}
 j_\ell (k \bar \chi)   \; , \label{deltam} \\
 \nonumber\\
&&{\mathcal S}_{\ell}^{[i] \p_\|^2 E'}(k)= (4 \pi) i^{\ell}  \int {\rm d} \bar \chi ~ { \mathcal W}^{[i]}( \bar\chi)\left[- {f(\eta) \over k^2}  {D(\eta) \over D(\eta_0)}\right] \left[ \frac{\partial^2}{\partial \bar \chi^2}  j_{\ell} ( k \bar \chi) \right] \;,  \label{Kaiser}  \\
 \nonumber\\
&&{\mathcal S}_{\ell }^{[i] \partial_\| E'}(k)=  (4 \pi) i^{\ell}  \int {\rm d} \bar \chi ~ { \mathcal W}^{[i]}( \bar\chi) \left[b^{[i]}_e(\eta) - 2 - {\cH'(\eta)\over \cH^2(\eta)} \right] \left[{\cH(\eta) f(\eta) \over k^2} {D(\eta) \over D(\eta_0)}\right]  \left[ \frac{\partial}{\partial \bar \chi}  j_\ell ( k \bar \chi) \right]\;,  \label{doppler}  \\
 \nonumber\\
&&{\mathcal S}_{\ell}^{[i] E''   }(k)= (4 \pi) i^{\ell}  \int {\rm d} \bar \chi ~ { \mathcal W}^{[i]}( \bar\chi) \left[b^{[i]}_e(\eta) - 2 - {\cH'(\eta)\over \cH^2(\eta)} \right]\left[{\cH^2(\eta)\Big(\frac{3}{2} \Omega_m (\eta) -f(\eta)\Big)\over k^2} {D(\eta) \over D(\eta_0)}\right]  j_{\ell} ( k \bar \chi)   \;,  \label{epp}  \\
 \nonumber\\
&&{\mathcal S}_{\ell }^{[i]  (\partial_\| E')_\ro}(k)=  -{(4 \pi)H_0f_0  i^{\ell}\over 3k}  \delta^K_{\ell_1} \int {\rm d} \bar \chi ~ { \mathcal W}^{[i]}( \bar\chi) \left[b^{[i]}_e(\eta) - 2 - {\cH'(\eta)\over \cH^2(\eta)} \right]   \;,  \label{dopplertoday}  \\
 \nonumber\\
&&{\mathcal S}_{\ell }^{[i] (E'')_\ro }(k)=  -{(4\pi) H_0^2 i^{\ell}\over k^2}\Big(\frac{3}{2} \Omega_{\rm m 0}-f_0\Big) \delta^K_{\ell 0}  \int {\rm d} \bar \chi ~ { \mathcal W}^{[i]}( \bar\chi) \left[b^{[i]}_e(\eta) - 2 - {\cH'(\eta)\over \cH^2(\eta)} \right]  \;,   \label{epptoday} \\
 \nonumber\\
&&{\mathcal S}_{\ell }^{[i]\int E''' }(k)=(4\pi) i^{\ell}  \int {\rm d} \bar \chi ~ { \mathcal W}^{[i]}( \bar\chi)   \left[b^{[i]}_e(\eta) - 2 - {\cH'(\eta)\over \cH^2(\eta)} \right]   \int_0^{\bar \chi} {\rm d} \tilde \chi ~\left[  3   \tilde{\cH}^3 \tilde{\Omega}_{\rm m} \left(\tilde f-1\right)k^{-2}  {D(\tilde \eta) \over D(\eta_0)}\right]  j_{\ell} ( k \tilde \chi)\;.
\label{isw}
\end{eqnarray}

In Appendix \ref{NEWdeltaGW} we will give a alternative definition of the Gravitational wave background anisotropy. As mentioned above, this is reminiscent of the CIB case, where projection effects provide also a large contribution to the signal (see~\cite{Tucci:2016hng}, where they are called GR corrections).

\section{Results}
\label{sec:results}
Here we compute the effect of projection effects on the inferred energy density, through the correlation function obtained analytically in the previous section. To study the importance of these projection effects, we consider a toy-model case: the ASGWB generated by black hole mergers in the frequency range of LIGO-Virgo and Einstein Telescope. Given the similarities between this formalism and the one used to compute the galaxy number counts angular power spectrum, we have modified the public code \texttt{CLASS} to compute the ASGWB anisotropies angular power spectrum. The details of the code will be presented in a companion paper in preparation~\cite{SGWB-II}.

Given that only unresolved sources contribute to the SGWB, the merger rate of black hole binaries has to be corrected with the detector efficiency. In particular, we assume a network of detectors composed by LIGO (Hanford and Livingston) and Virgo. The merger rate and detectability of merging events have been computed following the prescriptions of Ref.~\cite{Regimbau:2016ike}, while the GW waveform is computed as in Ref.~\cite{ajith:gwswaveform}, considering also the source orientations. In the following we consider the GWs emission in the~$f_\ro =50\ \mathrm{Hz}$ and~$f_\ro=200\ \mathrm{Hz}$ channels, assuming that all black hole binaries have members with masses~$\left(M_\mathrm{BH,1},M_\mathrm{BH,2}\right)=(15.0,15.0)$ and zero spin~$\left(\chi_1,\chi_2\right)=(0,0)$.

On the cosmological side, we compute the halo bias using the fitting formula calibrated on numerical simulations provided in Ref.~\cite{tinker:halobias}. The evolution bias is computed using the halo number density distribution of Ref.~\cite{tinker:halomassfunction}, also calibrated on numerical simulations. For simplicity, in the following we assume that all the events come from halos with mass~$M_\mathrm{halo}=10^{12}\ M_\odot$.

\begin{figure}[t!]
\centerline{
\includegraphics[width=1.0\textwidth]{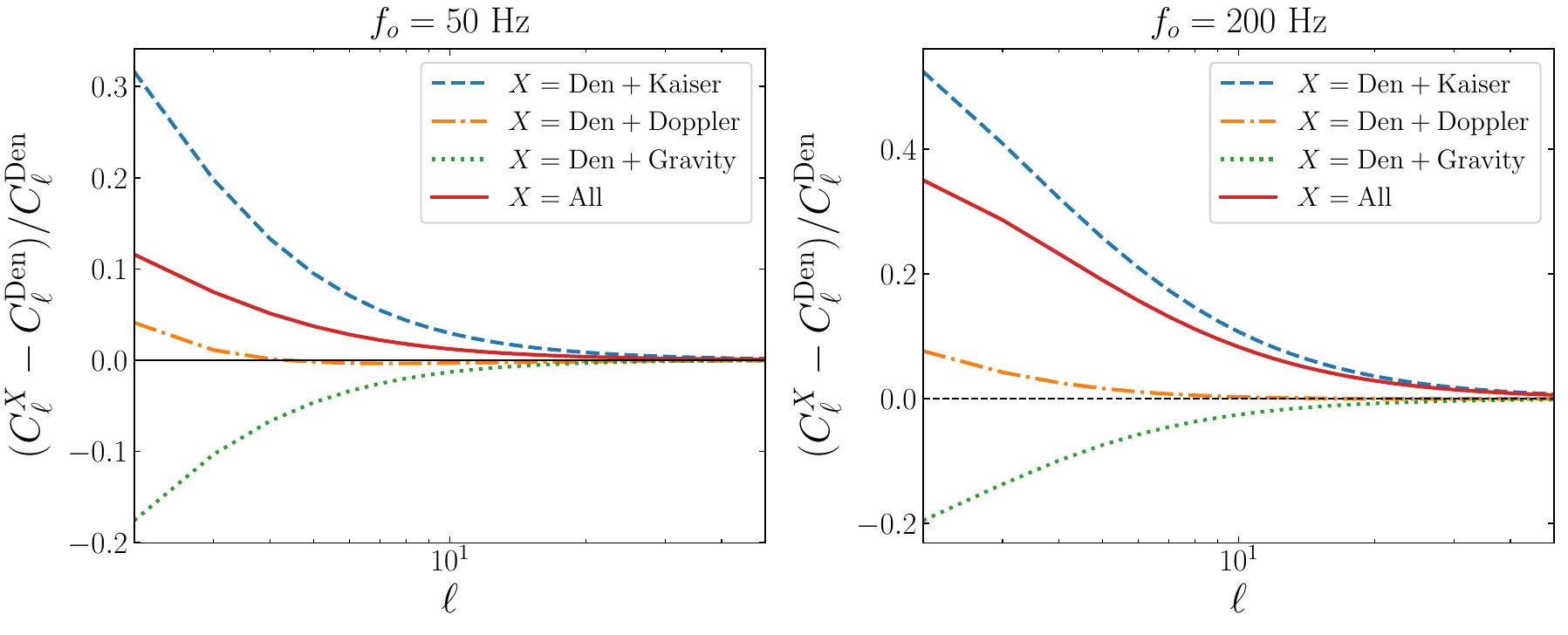}}
\caption{\it Relative contribution of different projection effects on the angular power spectra, for the~$50\ \mathrm{Hz}$ (\textit{left panel}) and~$200\ \mathrm{Hz}$ (\textit{right panel}) channels. In both cases we assume that all the events come from the merger of two 15 $M_\odot$ BHs with no spin, in $M_\mathrm{halo} = 10^{12}\ M_\odot$ dark matter halos, and we consider a second generation network of detectors.
The solid thin line at zero indicates intrinsic clustering; the blue dashed line shows the contribution of the Kaiser term, the dot-dashed orange is the Doppler effect, the dotted green shows the contribution of gravitational potential terms, and the red solid line shows the effect of all projection effects combined.
}
\label{fig:cls_contributions}
\end{figure}

In Figure~\ref{fig:cls_contributions} we report the relative difference of the angular power spectra including different contributions with respect to the angular power spectrum obtained using only density contributions (here for simplicity we are neglecting Equations~\eqref{dopplertoday} and~\eqref{epptoday}, evaluated at observer position).
The two panels show results for two different frequency channels, $50\ \mathrm{Hz}$ and $200\ \mathrm{Hz}$.
The solid thin line at zero indicates the intrinsic clustering; the blue dashed line shows the contribution of the Kaiser term, the dot-dashed orange is the Doppler effect, the dotted green shows the contribution of gravitational potential terms, and the red solid line shows the effect of all projection effects combined.
We showed relative contributions to both emphasize the effect of different terms and to isolate this from astrophysical and instrumental uncertainties.
The results presented show that the contribution of different effects: (i) is larger at the lowest angular multipoles; (ii) depends on the frequency of the signal measured.
By looking at the relative amplitude of the different effects, we can clearly see that they are all of the same order of magnitude, with the Kaiser term being, in our toy-model, the most important at all scales.
At the largest scales, Kaiser, Doppler, gravitational potentials terms contribute up to a few tens of percent of the total power spectrum amplitude.

In fact, the relative importance of different contributions depends on the choice of sources and detectors, i.e., on the~${ \mathcal W}^{[i]}$ function which has the role of a weight in Equations~\eqref{deltam}-\eqref{isw}.
This means that the relative importance of different effects depends on the assumed case, detector specifications, and astrophysical models.
A detailed investigation of these dependencies and their effect on cosmological and astrophysical model measurements will be presented in the follow-up paper~\cite{SGWB-II}.


\section{Conclusions}
\label{sec:conclusion}
The  detection and characterization of the astrophysical gravitational wave background will be another milestone in the GW community.  It will represent a further step toward studying astrophysical properties of black holes, cosmological model tests, and be a crucial part in the search for a gravitational wave background of cosmological origin, which would shed light on the physics of the early universe.
Many efforts have been already made to study the ASGWB  from analytical, numerical and map-making points of view, and the importance of
a precise understanding of both the astrophysical dependencies and propagation effects have been studied.
Such a process requires the development of consistent tools which can be directly compared to observations.

In this paper we compute the {\it observed} gravitational wave power spectrum, i.e., we use gauge invariant quantities accounting for all the effects intervening between the source and the observer.
We use the ``Cosmic Rulers'' formalism and we have taken into account all the corrections (at linear order) to the GW energy density.
The signal we are after is the stochastic superposition of unresolved astrophysical sources; it is generated by events that will be in principle resolved by higher precision instruments. Thus, this signal depends upon the instrument response, the survey strategy, and is affected by projection effects. It is analogous to other astrophysical backgrounds, and by its nature different from the primordial, inflationary generated, GW background.
It is therefore crucial to analyze (and before that, theoretically model) the signal, including all intervening effects and in the appropriate frame.
Mapping our perturbed quantity in the observed frame we are able to pick up information on the astrophysical properties of the GW sources and to obtain corrections due to GW propagation in an inhomogeneous universe. We considered two categories of sources: (I) events with short emission, e.g.,~merging binary sources and SNe explosions; (II) inspiralling binary sources which have not merged.
We derive expressions for all the (linear) contributions and investigate their relative contributions to the {\it observed} ASGWB.
Our results show that the contribution of different effects is larger at the lowest angular multipoles, and at the largest scales, Kaiser, Doppler, gravitational potentials terms contribute up to a few tens of percent of the total power spectrum amplitude, and their importance is also frequency-dependent. Among these, the Kaiser term plays a relevant role, as in the case of other astrophysical backgrounds like the CIB~\cite{Tucci:2016hng}, radio-continuum and intensity mapping, as shown in \cite{Hall:2012wd, Alonso:2015sfa}. From our plots we see that the Kaiser term is dominant on large angular scales, while becoming negligible on small scales. Finally, let us mention that when the integration along the line of sight is performed, one should also consider the normalized selection window function, whose form depends, besides redshift,  on the sensitivity/characteristics of the detector (e.g. interferometers in the case of GW).

Therefore, with the caveat of astrophysical and instrumental dependencies, which will need to be investigated in more detail, we see an indication that projection effects will need to be included in all theoretical modeling of the ASGWB.

A follow-up paper in preparation~\cite{SGWB-II} will describe the numerical implementation of this formalism, a more advanced astrophysical modeling and realistic black hole mass functions, and present a detailed analysis in order to estimate the astrophysical parameters derived for astrophysical background sources.

\section*{Acknowledgements}
We thank Vincent Desjacques, Ely Kovetz, Licia Verde, Zvonimir Vlah for useful discussions.
DB, AnRi, NBe, SM, AlRa would like to thank Giulia Cusin for useful discussions at the LISA CosWG meeting in Padova. DB and SM acknowledge partial financial support by ASI Grant No. 2016-24-H.0. NBe is supported by the Spanish MINECO under grant BES-2015-073372.
 ACJ is supported by King's College London through a Graduate Teaching Scholarship. MS is supported in part by the Science and Technology Facility Council (STFC), United Kingdom, under the research grant ST/P000258/1.

\appendix
\section{}\label{A}
In our analysis we often prefer to work with projected quantities, in order to identify the various contributes along and perpendicular with respect to the line of sight. Imagine to consider a three dimensional Cartesian reference frame in which  the $z$ axis connects us (the observers) to the starting point of the GW we detect, thus defining the so-called line of sight which is basically identified with the direction $\bf{\hat{n}}$ and the $x-y$ plane is perpendicular to the line of sight and passes through the source position. With this in mind we  can define and projected along or perpendicular to the line of sight.  In this way the quantities we work with  as derivative operators, vectors and tensors, can be defined as
\begin{eqnarray}
\label{Projection}
A_{\parallel} = n^{i} n^{j} A_{ij } \;, \quad
B_{\perp}^i =  \Perp^{ij} B_j = B^i -n^i B_{\parallel}\;,
\end{eqnarray}
where $\Perp^i_{j}= \delta^{i}_j-n^in_j$. The same can be done with derivatives which become
\begin{eqnarray}
\label{Projection2}
\bar \p_\parallel = n^i  {\p \over   \p \bar x^i} \;,  \quad \quad \bar \p^2_\parallel  = \bar \p_\parallel \bar \p_\parallel\;,\quad \quad  \bar \p_{\perp i} =  \Perp^j_i \bar \p_j=   {\p \over   \p \bar x^i} -  n_i \bar \p_\parallel\;, \quad \quad   {\p n^j \over   \p \bar x^i}  &=& \frac{1}{\bar \chi}\Perp_i^j
,\quad \quad \frac{\ud}{\ud \bar \chi} \p_{\perp}^i = \bar \p_{\perp}^i  \frac{\ud}{\ud \bar \chi}-\frac{1}{\bar \chi} \p_{\perp}^i
\;,
\end{eqnarray}
and we have
\begin{eqnarray}
\label{Projection3}
 {\p B^i \over   \p \bar x^j}  &=& n^i n_j \bar \p_{\parallel} B_{\parallel} + n^i \bar \p_{\perp j} B_{\parallel} +\bar \p_{\perp j} B_{\perp}^i+ n_j \bar \p_{\parallel} B_{\perp}^i +\frac{1}{\bar \chi} \Perp^i_j B_{\parallel}\;,\nonumber \\
\quad \quad  \bar \nabla^2_\perp &=&\bar  \p_{\perp i} \bar \p_\perp^i = \delta^{ij} {\p \over   \p \bar x^i}  {\p \over   \p \bar x^j} - \bar \partial_\parallel^2 - \frac{2}{\bar \chi} \bar \p_\parallel\;.
\end{eqnarray}

\section{Proof of $\delta x^{0}_\ro$ and $\delta x^{i}_\ro$}\label{B}

We compute here explicitly $\delta x^{0}_\ro$ and $\delta x^{i}_\ro$; which arise from the fact that the physical coordinate time $t_0=t(\eta=\eta_0)=t_{\rm in}+\int_{\eta_{\rm in}}^{\eta_0} a(\tilde \eta) \ud\tilde \eta$ does not coincide with the proper time of the observer $\T_0$ in an inhomogeneous universe. We have
\be
\frac{\ud  x^\mu}{\ud \T}=u^\mu={E_{\hat 0}^\mu\over a}\;,
\ee
then for $\mu=0$ and considering the physical coordinate $\ud t=a(\eta) \ud \eta$ at the observer we have
\[t_0-t_{\rm in}=\T_0-\T_{\rm in}+\int_{\T_{\rm in}}^{\T_0} E_{\hat 0}^0 \; \ud\T\;.\]
Now $t_0=\bar t_0 + \delta t_\ro$,  where $\bar t_0$ is the time coordinate of the observer and, therefore, it has to coincide with the proper time, i.e.
\[ \bar t_0-t_{\rm in }= \T_0-\T_{\rm in} \]
and $\delta t_\ro=\bar a_\ro \delta \eta_\ro= \delta \eta_\ro$ we have
\be
\delta x^{0}_\ro=\delta \eta_\ro= \int_{\bar \eta_{\rm in}}^{\bar \eta_0} \bar a \; E_{\hat 0}^0 \ud \tilde \eta=- \int_{\bar \eta_{\rm in}}^{\bar \eta_0} \bar a(\tilde \eta) \; A(\tilde \eta, {\bf 0})\; \ud \tilde \eta
\ee
Finally, for $\mu=i$
\bea
\delta x^{i}_\ro &=&  \int_{\T_{\rm in}}^{\T_\ro}  \delta u^i \ud \tilde \T=  \int_{\bar \eta_{\rm in}}^{\bar \eta_0} v^i (\tilde \eta, {\bf 0})\; \ud \tilde \eta\;.
\eea
As reported in the main text.

\section{Poisson Gauge}\label{C}

In this section we write all GR effects in Poisson Gauge (P).
Starting from  Eq.~\eqref{metric} in P gauge we have $A^{\rm (P)}=\Psi$, $B^{\rm (P)}_i=0$, $h^{\rm (P)}_{ij}=-2\Phi\delta_{ij}$, $F^{\rm (P)}_{ij}=0$, where $\Psi$ and $\Phi$ are the Bardeen potentials. By assuming the concordance background model and,  at first order, neglecting the anisotropic stress, we have $\Psi=\Phi$ and
\be
 \label{Poiss-metric}
 {\rm d} s^2 = a(\eta)^2\left[-\left(1 + 2\Phi\right)  {\rm d}  \eta^2+ \delta_{ij} \left(1 -2\Phi \right)  {\rm d}   x^i  {\rm d}  x^j\right] .\\
\ee
 In this gauge $v_\parallel=n^i v_i=\hat{\mathbf{n}}\cdot\mathbf{v}$ (where $v^{i  }= \partial^i v$),

 $\delta a_o = - H_0 \int_{\bar \eta_{\rm in}}^{\bar \eta_0} \bar a(\tilde \eta) \Phi(\tilde \eta, {\bf 0})\; \ud \tilde \eta$, the
 (Shapiro) time-delay (STD) term is
\begin{equation}
\label{Poiss-s-1}
 T^{\rm (P)} =- 2 \int_0^{\bar \chi} {\rm d} \tilde \chi \Phi \;;
\end{equation} and
\begin{equation}\label{Poiss-iota}
I^{\rm (P)}   =  - \int_0^{\bar \chi} {\rm d} \tilde \chi \, \Phi {'}  \;,
\end{equation}
 is the integrated Sachs-Wolfe (ISW) effect. For dark matter particles in general relativity we have $\mathbf{v}' + \mathcal{H}\mathbf{v} + \nabla\Phi = 0$
and the GWs overdensity in Poisson gauge is written as
\be
\delta^{[i] {\rm (P)}} =  \delta^{[i] {\rm (SC)}} - b_e \cH v+ 3 \cH v = b^{[i]} (\eta) \delta^{[i]}_{\rm m} - b_e \cH v+ 3 \cH v
\;,
\ee
where we used Eq. (\ref{def-bias}).

\begin{eqnarray}
\label{DeltaOmegaPoisson}
 \Delta \Omega_{\rm GW}({\bf n}) &=& \frac{ f_\ro}{\rho_c}  \sum_{[i]} \int {N^{[i]} [z, f_\ro (1+z)] \over (1+z)}   \nonumber \\
 &\times&  \Bigg\{ b^{[i]} \delta^{[i]}_{\rm m}  + \left(b^{[i]}_{e} - 2 - \frac{\mathcal{H}'}{\mathcal{H}^2}\right) \hat{\mathbf{n}}\cdot\mathbf{v} - \frac{1}{\mathcal{H}} \partial_\parallel(\hat{\mathbf{n}}\cdot\mathbf{v}) - (b^{[i]}_{e}-3) \mathcal{H} v + \nonumber \\
&\qquad& + \left(3 - b^{[i]}_{e} + \frac{\mathcal{H}'}{\mathcal{H}^2}\right) \Phi + \frac{1}{\mathcal{H}}\Phi' +2 \left(2 - b^{[i]}_\mathrm{e} + \frac{\mathcal{H}'}{\mathcal{H}^2}\right) \int_0^{\bar \chi} d\tilde \chi \Phi' + \nonumber\\
&\qquad& + \left(b^{[i]}_\mathrm{e} - 2 - \frac{\mathcal{H}'}{\mathcal{H}^2}\right) \left[-\mathcal{H}_0 \left(\int_{\bar \eta_{\rm in}}^{\bar \eta_0} d\tilde \eta {\Phi(\tilde \eta, {\bf 0} )\over (1+z(\tilde \eta ))}\right) + \Phi_\ro -  \left(\hat{\mathbf{n}}\cdot\mathbf{v} \right)_\ro \right]\Bigg\} \; \ud  \bar \chi   .
\end{eqnarray}

\section{Alternative definition of the gravitational wave background anisotropy} \label{NEWdeltaGW}
Rewriting the background  energy density in the following way
\be
{ \bar \Omega_{\rm GW}\over 4\pi}  = \sum_{[i]}{ \bar \Omega^{[i]}_{\rm GW}\over 4\pi} \;,
\ee
where
\be
{ \bar \Omega^{[i]}_{\rm GW}\over 4\pi} = \frac{ f_\ro}{\rho_{\rm c}}  \int  {N^{[i]} (z, f_\re) \over  (1+z)}  \; \ud  \bar \chi \;,
\ee
we are able to define a new quantity, i.e. the GW energy-density overdensity,
\be
\Delta_{\rm GW}= \frac{\Delta  \Omega_{\rm GW}}{\bar \Omega_{\rm GW}/4\pi}=
\sum_{[i]}
f^{[i]}_{\rm GW} \Delta^{[i]}_{\rm GW}\;,
\ee
where
\be
f^{[i]}_{\rm GW} \equiv \frac{\bar \Omega^{[i]}_{\rm GW}}{\bar \Omega_{\rm GW}}
\ee
is the weight of the relative contribution of the sources which is bounded to be~$f^{[i]}_{\rm GW} \in [0,1]$. Here  $\Delta^{[i]}_{\rm GW}$ is  the GW energy-density contrast for each contribution. Note that, using this new definition, it is possible to describe quickly both the ASGW and CSWG, and compute the angular power spectrum of the GW energy density contrast
\begin{align}
\label{eq:clgg2}
\D^{\rm GW}_{\ell}=\sum_{i,j;\alpha,\beta} \D^{[ij]\alpha\beta}_\ell\,
\end{align}
where
 \be
 \D^{[ij]\alpha \beta}_\ell = f^{[i]}_{\rm GW} f^{[j]}_{\rm GW} \sum_{\ell=- m}^{\ell=m}\frac{\left\langle {\beta}^{[i]\alpha*}_{\ell m}~
{\beta}^{[j]\beta}_{\ell m} \right\rangle}{2\ell+1} = f^{[i]}_{\rm GW} f^{[j]}_{\rm GW} \int {k^2 \ud k \over(2\pi)^3} \tilde{\mathcal S}^{[i]\alpha*}_{\ell} (k) \tilde {\mathcal S}^{[j]\beta}_{ \ell}(k) P_{\rm m}(k) \,  \, ,
\ee
with
\be
{\beta^{[i]\alpha}}_{\ell m} =  \int  \frac{{\rm d}^3  { \bf k}}{(2\pi)^3}\, Y^*_{\ell m}(\hat{\bf k})   {\tilde {\mathcal S}}^{[i] \alpha}_{\ell}(k)  \delta_{\rm m}({ \bf k},\eta_0)
\ee
and
\begin{eqnarray}
{\tilde {\mathcal S}}^{[i]\alpha}(k) \equiv 4 \pi i^{\ell} \int {\rm d}{\bar \chi}   ~ {\tilde{ \mathcal W}}^{[i]}( \bar\chi)\int_0^{\bar \chi} {\rm d} \tilde \chi ~\left[\mathbb{ W}^{\alpha} \left(\bar \chi, \tilde \chi, \eta, \tilde \eta, \frac{\partial}{\partial \tilde \chi}, \frac{\partial}{\partial \tilde \eta}\right)  {\Upsilon}^{\alpha}( {\bf k},\tilde \eta) j_{\ell} ( k \tilde \chi)\right].
\end{eqnarray}
Here we have defined  a new weight function
\be
{\tilde{ \mathcal W}}^{[i]}( \bar\chi)=\frac{ f_\ro}{\rho_c}  \frac{4\pi }{\bar \Omega^{[i]}_{\rm GW}} {N^{[i]} [ z, f_\ro (1+z)] \over  (1+z)}  \;.
\ee
Finally,  $\tilde{ \mathcal S}_{\ell}^a(k)$ functions read as
\begin{eqnarray}
&&\tilde {\mathcal S}_{\ell }^{[i] \delta_{\rm m}^{\rm (SC)} }(k)= (4 \pi) i^{\ell} \int {\rm d} \bar \chi ~\tilde { \mathcal W}^{[i]}( \bar\chi)  b_{\mathrm{gw}}^{[i]} (\eta) {D(\eta) \over D(\eta_0)}
 j_\ell (k \bar \chi)   \; , \nonumber \\
 \nonumber\\
&&\tilde {\mathcal S}_{\ell}^{[i] \p_\|^2 E'}(k)= (4 \pi) i^{\ell}  \int {\rm d} \bar \chi ~ \tilde{ \mathcal W}^{[i]}( \bar\chi)\left[- {f(\eta) \over k^2}  {D(\eta) \over D(\eta_0)}\right] \left[ \frac{\partial^2}{\partial \bar \chi^2}  j_{\ell} ( k \bar \chi) \right] \;, \nonumber   \\
 \nonumber\\
&&\tilde {\mathcal S}_{\ell }^{[i] \partial_\| E'}(k)=  (4 \pi) i^{\ell}  \int {\rm d} \bar \chi ~ \tilde{ \mathcal W}^{[i]}( \bar\chi) \left[b^{[i]}_e(\eta) - 2 - {\cH'(\eta)\over \cH^2(\eta)} \right] \left[{\cH(\eta) f(\eta) \over k^2} {D(\eta) \over D(\eta_0)}\right]  \left[ \frac{\partial}{\partial \bar \chi}  j_\ell ( k \bar \chi) \right]\;,   \nonumber  \\
 \nonumber\\
&&\tilde {\mathcal S}_{\ell}^{[i] E''   }(k)= (4 \pi) i^{\ell}  \int {\rm d} \bar \chi ~ \tilde{ \mathcal W}^{[i]}( \bar\chi) \left[b^{[i]}_e(\eta) - 2 - {\cH'(\eta)\over \cH^2(\eta)} \right]\left[{\cH^2(\eta)\Big(\frac{3}{2} \Omega_m (\eta) -f(\eta)\Big)\over k^2} {D(\eta) \over D(\eta_0)}\right]  j_{\ell} ( k \bar \chi)   \;, \nonumber \\
 \nonumber\\
&&\tilde {\mathcal S}_{\ell }^{[i]  (\partial_\| E')_\ro}(k)=  -{(4 \pi)H_0f_0  i^{\ell}\over 3k}  \delta^K_{\ell_1} \int {\rm d} \bar \chi ~ \tilde { \mathcal W}^{[i]}( \bar\chi) \left[b^{[i]}_e(\eta) - 2 - {\cH'(\eta)\over \cH^2(\eta)} \right]   \;,\nonumber   \\
 \nonumber\\
&&\tilde {\mathcal S}_{\ell }^{[i] (E'')_\ro }(k)=  -{(4\pi) H_0^2 i^{\ell}\over k^2}\Big(\frac{3}{2} \Omega_{\rm m 0}-f_0\Big) \delta^K_{\ell 0}  \int {\rm d} \bar \chi ~ \tilde { \mathcal W}^{[i]}( \bar\chi) \left[b^{[i]}_e(\eta) - 2 - {\cH'(\eta)\over \cH^2(\eta)} \right]  \;,  \nonumber \\
 \nonumber\\
&&\tilde {\mathcal S}_{\ell }^{[i]\int E''' }(k)=(4\pi) i^{\ell}  \int {\rm d} \bar \chi ~\tilde  { \mathcal W}^{[i]}( \bar\chi)   \left[b^{[i]}_e(\eta) - 2 - {\cH'(\eta)\over \cH^2(\eta)} \right]   \int_0^{\bar \chi} {\rm d} \tilde \chi ~\left[  3   \tilde{\cH}^3 \tilde{\Omega}_{\rm m} \left(\tilde f-1\right)k^{-2}  {D(\tilde \eta) \over D(\eta_0)}\right]  j_{\ell} ( k \tilde \chi)\;.
\end{eqnarray}

\bibliography{RefAGWB}

\begin{thebibliography}{58}
\expandafter\ifx\csname natexlab\endcsname\relax\def\natexlab#1{#1}\fi
\expandafter\ifx\csname bibnamefont\endcsname\relax
  \def\bibnamefont#1{#1}\fi
\expandafter\ifx\csname bibfnamefont\endcsname\relax
  \def\bibfnamefont#1{#1}\fi
\expandafter\ifx\csname citenamefont\endcsname\relax
  \def\citenamefont#1{#1}\fi
\expandafter\ifx\csname url\endcsname\relax
  \def\url#1{\texttt{#1}}\fi
\expandafter\ifx\csname urlprefix\endcsname\relax\def\urlprefix{URL }\fi
\providecommand{\bibinfo}[2]{#2}
\providecommand{\eprint}[2][]{\url{#2}}

\bibitem[{\citenamefont{Abbott et~al.}(2018)}]{Aasi:2013wya}
\bibinfo{author}{\bibfnamefont{B.~P.} \bibnamefont{Abbott}}
  \bibnamefont{et~al.} (\bibinfo{collaboration}{KAGRA, LIGO Scientific,
  VIRGO}), \bibinfo{journal}{Living Rev. Rel.} \textbf{\bibinfo{volume}{21}},
  \bibinfo{pages}{3} (\bibinfo{year}{2018}), \eprint{1304.0670}.

\bibitem[{\citenamefont{Akrami et~al.}(2018)}]{Akrami:2018odb}
\bibinfo{author}{\bibfnamefont{Y.}~\bibnamefont{Akrami}} \bibnamefont{et~al.}
  (\bibinfo{collaboration}{Planck}) (\bibinfo{year}{2018}),
  \eprint{1807.06211}.

\bibitem[{\citenamefont{Maggiore}(2000)}]{Maggiore:1999vm}
\bibinfo{author}{\bibfnamefont{M.}~\bibnamefont{Maggiore}},
  \bibinfo{journal}{Phys. Rept.} \textbf{\bibinfo{volume}{331}},
  \bibinfo{pages}{283} (\bibinfo{year}{2000}), \eprint{gr-qc/9909001}.

\bibitem[{\citenamefont{Guzzetti et~al.}(2016)\citenamefont{Guzzetti, Bartolo,
  Liguori, and Matarrese}}]{Guzzetti:2016mkm}
\bibinfo{author}{\bibfnamefont{M.~C.} \bibnamefont{Guzzetti}},
  \bibinfo{author}{\bibfnamefont{N.}~\bibnamefont{Bartolo}},
  \bibinfo{author}{\bibfnamefont{M.}~\bibnamefont{Liguori}}, \bibnamefont{and}
  \bibinfo{author}{\bibfnamefont{S.}~\bibnamefont{Matarrese}},
  \bibinfo{journal}{Riv. Nuovo Cim.} \textbf{\bibinfo{volume}{39}},
  \bibinfo{pages}{399} (\bibinfo{year}{2016}), \eprint{1605.01615}.

\bibitem[{\citenamefont{Bartolo et~al.}(2016)}]{Bartolo:2016ami}
\bibinfo{author}{\bibfnamefont{N.}~\bibnamefont{Bartolo}} \bibnamefont{et~al.},
  \bibinfo{journal}{JCAP} \textbf{\bibinfo{volume}{1612}}, \bibinfo{pages}{026}
  (\bibinfo{year}{2016}), \eprint{1610.06481}.

\bibitem[{\citenamefont{Caprini and Figueroa}(2018)}]{Caprini:2018mtu}
\bibinfo{author}{\bibfnamefont{C.}~\bibnamefont{Caprini}} \bibnamefont{and}
  \bibinfo{author}{\bibfnamefont{D.~G.} \bibnamefont{Figueroa}},
  \bibinfo{journal}{Class. Quant. Grav.} \textbf{\bibinfo{volume}{35}},
  \bibinfo{pages}{163001} (\bibinfo{year}{2018}), \eprint{1801.04268}.

\bibitem[{\citenamefont{Regimbau}(2011)}]{Regimbau:2011rp}
\bibinfo{author}{\bibfnamefont{T.}~\bibnamefont{Regimbau}},
  \bibinfo{journal}{Res. Astron. Astrophys.} \textbf{\bibinfo{volume}{11}},
  \bibinfo{pages}{369} (\bibinfo{year}{2011}), \eprint{1101.2762}.

\bibitem[{\citenamefont{Romano and Cornish}(2017)}]{Romano:2016dpx}
\bibinfo{author}{\bibfnamefont{J.~D.} \bibnamefont{Romano}} \bibnamefont{and}
  \bibinfo{author}{\bibfnamefont{N.~J.} \bibnamefont{Cornish}},
  \bibinfo{journal}{Living Rev. Rel.} \textbf{\bibinfo{volume}{20}},
  \bibinfo{pages}{2} (\bibinfo{year}{2017}), \eprint{1608.06889}.

\bibitem[{\citenamefont{Bartolo et~al.}(2019)\citenamefont{Bartolo, Bertacca,
  Matarrese, Peloso, Ricciardone, Riotto, and Tasinato}}]{Bartolo:2019oiq}
\bibinfo{author}{\bibfnamefont{N.}~\bibnamefont{Bartolo}},
  \bibinfo{author}{\bibfnamefont{D.}~\bibnamefont{Bertacca}},
  \bibinfo{author}{\bibfnamefont{S.}~\bibnamefont{Matarrese}},
  \bibinfo{author}{\bibfnamefont{M.}~\bibnamefont{Peloso}},
  \bibinfo{author}{\bibfnamefont{A.}~\bibnamefont{Ricciardone}},
  \bibinfo{author}{\bibfnamefont{A.}~\bibnamefont{Riotto}}, \bibnamefont{and}
  \bibinfo{author}{\bibfnamefont{G.}~\bibnamefont{Tasinato}}
  (\bibinfo{year}{2019}), \eprint{1908.00527}.

\bibitem[{\citenamefont{Shannon et~al.}(2013)}]{Shannon:2013wma}
\bibinfo{author}{\bibfnamefont{R.~M.} \bibnamefont{Shannon}}
  \bibnamefont{et~al.}, \bibinfo{journal}{Science}
  \textbf{\bibinfo{volume}{342}}, \bibinfo{pages}{334} (\bibinfo{year}{2013}),
  \eprint{1310.4569}.

\bibitem[{\citenamefont{Abbott et~al.}(2019)}]{LIGOScientific:2019vic}
\bibinfo{author}{\bibfnamefont{B.~P.} \bibnamefont{Abbott}}
  \bibnamefont{et~al.} (\bibinfo{collaboration}{LIGO Scientific, Virgo}),
  \bibinfo{journal}{Phys. Rev.} \textbf{\bibinfo{volume}{D100}},
  \bibinfo{pages}{061101} (\bibinfo{year}{2019}), \eprint{1903.02886}.

\bibitem[{\citenamefont{Taylor et~al.}(2015)}]{Taylor:2015udp}
\bibinfo{author}{\bibfnamefont{S.~R.} \bibnamefont{Taylor}}
  \bibnamefont{et~al.}, \bibinfo{journal}{Phys. Rev. Lett.}
  \textbf{\bibinfo{volume}{115}}, \bibinfo{pages}{041101}
  (\bibinfo{year}{2015}), \eprint{1506.08817}.

\bibitem[{\citenamefont{Alba and Maldacena}(2016)}]{Alba:2015cms}
\bibinfo{author}{\bibfnamefont{V.}~\bibnamefont{Alba}} \bibnamefont{and}
  \bibinfo{author}{\bibfnamefont{J.}~\bibnamefont{Maldacena}},
  \bibinfo{journal}{JHEP} \textbf{\bibinfo{volume}{03}}, \bibinfo{pages}{115}
  (\bibinfo{year}{2016}), \eprint{1512.01531}.

\bibitem[{\citenamefont{Contaldi}(2017)}]{Contaldi:2016koz}
\bibinfo{author}{\bibfnamefont{C.~R.} \bibnamefont{Contaldi}},
  \bibinfo{journal}{Phys. Lett.} \textbf{\bibinfo{volume}{B771}},
  \bibinfo{pages}{9} (\bibinfo{year}{2017}), \eprint{1609.08168}.

\bibitem[{\citenamefont{Cusin et~al.}(2017)\citenamefont{Cusin, Pitrou, and
  Uzan}}]{Cusin:2017fwz}
\bibinfo{author}{\bibfnamefont{G.}~\bibnamefont{Cusin}},
  \bibinfo{author}{\bibfnamefont{C.}~\bibnamefont{Pitrou}}, \bibnamefont{and}
  \bibinfo{author}{\bibfnamefont{J.-P.} \bibnamefont{Uzan}},
  \bibinfo{journal}{Phys. Rev.} \textbf{\bibinfo{volume}{D96}},
  \bibinfo{pages}{103019} (\bibinfo{year}{2017}), \eprint{1704.06184}.

\bibitem[{\citenamefont{Cusin et~al.}(2018{\natexlab{a}})\citenamefont{Cusin,
  Dvorkin, Pitrou, and Uzan}}]{Cusin:2018rsq}
\bibinfo{author}{\bibfnamefont{G.}~\bibnamefont{Cusin}},
  \bibinfo{author}{\bibfnamefont{I.}~\bibnamefont{Dvorkin}},
  \bibinfo{author}{\bibfnamefont{C.}~\bibnamefont{Pitrou}}, \bibnamefont{and}
  \bibinfo{author}{\bibfnamefont{J.-P.} \bibnamefont{Uzan}},
  \bibinfo{journal}{Phys. Rev. Lett.} \textbf{\bibinfo{volume}{120}},
  \bibinfo{pages}{231101} (\bibinfo{year}{2018}{\natexlab{a}}),
  \eprint{1803.03236}.

\bibitem[{\citenamefont{Cusin et~al.}(2018{\natexlab{b}})\citenamefont{Cusin,
  Pitrou, and Uzan}}]{Cusin:2017mjm}
\bibinfo{author}{\bibfnamefont{G.}~\bibnamefont{Cusin}},
  \bibinfo{author}{\bibfnamefont{C.}~\bibnamefont{Pitrou}}, \bibnamefont{and}
  \bibinfo{author}{\bibfnamefont{J.-P.} \bibnamefont{Uzan}},
  \bibinfo{journal}{Phys. Rev.} \textbf{\bibinfo{volume}{D97}},
  \bibinfo{pages}{123527} (\bibinfo{year}{2018}{\natexlab{b}}),
  \eprint{1711.11345}.

\bibitem[{\citenamefont{Jenkins and Sakellariadou}(2018)}]{Jenkins:2018lvb}
\bibinfo{author}{\bibfnamefont{A.~C.} \bibnamefont{Jenkins}} \bibnamefont{and}
  \bibinfo{author}{\bibfnamefont{M.}~\bibnamefont{Sakellariadou}},
  \bibinfo{journal}{Phys. Rev.} \textbf{\bibinfo{volume}{D98}},
  \bibinfo{pages}{063509} (\bibinfo{year}{2018}), \eprint{1802.06046}.

\bibitem[{\citenamefont{Jenkins et~al.}(2018)\citenamefont{Jenkins,
  Sakellariadou, Regimbau, and Slezak}}]{Jenkins:2018uac}
\bibinfo{author}{\bibfnamefont{A.~C.} \bibnamefont{Jenkins}},
  \bibinfo{author}{\bibfnamefont{M.}~\bibnamefont{Sakellariadou}},
  \bibinfo{author}{\bibfnamefont{T.}~\bibnamefont{Regimbau}}, \bibnamefont{and}
  \bibinfo{author}{\bibfnamefont{E.}~\bibnamefont{Slezak}},
  \bibinfo{journal}{Phys. Rev.} \textbf{\bibinfo{volume}{D98}},
  \bibinfo{pages}{063501} (\bibinfo{year}{2018}), \eprint{1806.01718}.

\bibitem[{\citenamefont{Jenkins
  et~al.}(2019{\natexlab{a}})\citenamefont{Jenkins, O'Shaughnessy,
  Sakellariadou, and Wysocki}}]{Jenkins:2018kxc}
\bibinfo{author}{\bibfnamefont{A.~C.} \bibnamefont{Jenkins}},
  \bibinfo{author}{\bibfnamefont{R.}~\bibnamefont{O'Shaughnessy}},
  \bibinfo{author}{\bibfnamefont{M.}~\bibnamefont{Sakellariadou}},
  \bibnamefont{and} \bibinfo{author}{\bibfnamefont{D.}~\bibnamefont{Wysocki}},
  \bibinfo{journal}{Phys. Rev. Lett.} \textbf{\bibinfo{volume}{122}},
  \bibinfo{pages}{111101} (\bibinfo{year}{2019}{\natexlab{a}}),
  \eprint{1810.13435}.

\bibitem[{\citenamefont{Cusin et~al.}(2019{\natexlab{a}})\citenamefont{Cusin,
  Dvorkin, Pitrou, and Uzan}}]{Cusin:2019jpv}
\bibinfo{author}{\bibfnamefont{G.}~\bibnamefont{Cusin}},
  \bibinfo{author}{\bibfnamefont{I.}~\bibnamefont{Dvorkin}},
  \bibinfo{author}{\bibfnamefont{C.}~\bibnamefont{Pitrou}}, \bibnamefont{and}
  \bibinfo{author}{\bibfnamefont{J.-P.} \bibnamefont{Uzan}}
  (\bibinfo{year}{2019}{\natexlab{a}}), \eprint{1904.07797}.

\bibitem[{\citenamefont{Cusin et~al.}(2019{\natexlab{b}})\citenamefont{Cusin,
  Dvorkin, Pitrou, and Uzan}}]{Cusin:2019jhg}
\bibinfo{author}{\bibfnamefont{G.}~\bibnamefont{Cusin}},
  \bibinfo{author}{\bibfnamefont{I.}~\bibnamefont{Dvorkin}},
  \bibinfo{author}{\bibfnamefont{C.}~\bibnamefont{Pitrou}}, \bibnamefont{and}
  \bibinfo{author}{\bibfnamefont{J.-P.} \bibnamefont{Uzan}}
  (\bibinfo{year}{2019}{\natexlab{b}}), \eprint{1904.07757}.

\bibitem[{\citenamefont{Jenkins and Sakellariadou}(2019)}]{Jenkins:2019uzp}
\bibinfo{author}{\bibfnamefont{A.~C.} \bibnamefont{Jenkins}} \bibnamefont{and}
  \bibinfo{author}{\bibfnamefont{M.}~\bibnamefont{Sakellariadou}}
  (\bibinfo{year}{2019}), \eprint{1902.07719}.

\bibitem[{\citenamefont{Jenkins
  et~al.}(2019{\natexlab{b}})\citenamefont{Jenkins, Romano, and
  Sakellariadou}}]{Jenkins:2019nks}
\bibinfo{author}{\bibfnamefont{A.~C.} \bibnamefont{Jenkins}},
  \bibinfo{author}{\bibfnamefont{J.~D.} \bibnamefont{Romano}},
  \bibnamefont{and}
  \bibinfo{author}{\bibfnamefont{M.}~\bibnamefont{Sakellariadou}}
  (\bibinfo{year}{2019}{\natexlab{b}}), \eprint{1907.06642}.

\bibitem[{\citenamefont{Jeong et~al.}(2012)\citenamefont{Jeong, Schmidt, and
  Hirata}}]{Jeong:2011as}
\bibinfo{author}{\bibfnamefont{D.}~\bibnamefont{Jeong}},
  \bibinfo{author}{\bibfnamefont{F.}~\bibnamefont{Schmidt}}, \bibnamefont{and}
  \bibinfo{author}{\bibfnamefont{C.~M.} \bibnamefont{Hirata}},
  \bibinfo{journal}{Phys. Rev.} \textbf{\bibinfo{volume}{D85}},
  \bibinfo{pages}{023504} (\bibinfo{year}{2012}), \eprint{1107.5427}.

\bibitem[{\citenamefont{Schmidt and Jeong}(2012)}]{Schmidt:2012ne}
\bibinfo{author}{\bibfnamefont{F.}~\bibnamefont{Schmidt}} \bibnamefont{and}
  \bibinfo{author}{\bibfnamefont{D.}~\bibnamefont{Jeong}},
  \bibinfo{journal}{Phys. Rev.} \textbf{\bibinfo{volume}{D86}},
  \bibinfo{pages}{083527} (\bibinfo{year}{2012}), \eprint{1204.3625}.

\bibitem[{\citenamefont{Bertacca et~al.}(2018)\citenamefont{Bertacca,
  Raccanelli, Bartolo, and Matarrese}}]{Bertacca:2017vod}
\bibinfo{author}{\bibfnamefont{D.}~\bibnamefont{Bertacca}},
  \bibinfo{author}{\bibfnamefont{A.}~\bibnamefont{Raccanelli}},
  \bibinfo{author}{\bibfnamefont{N.}~\bibnamefont{Bartolo}}, \bibnamefont{and}
  \bibinfo{author}{\bibfnamefont{S.}~\bibnamefont{Matarrese}},
  \bibinfo{journal}{Phys. Dark Univ.} \textbf{\bibinfo{volume}{20}},
  \bibinfo{pages}{32} (\bibinfo{year}{2018}), \eprint{1702.01750}.

\bibitem[{\citenamefont{Ade et~al.}(2011)}]{PlanckCIB}
\bibinfo{author}{\bibfnamefont{P.~A.~R.} \bibnamefont{Ade}}
  \bibnamefont{et~al.} (\bibinfo{collaboration}{Planck}),
  \bibinfo{journal}{Astron. Astrophys.} \textbf{\bibinfo{volume}{536}},
  \bibinfo{pages}{A18} (\bibinfo{year}{2011}), \eprint{1101.2028}.

\bibitem[{\citenamefont{Tucci et~al.}(2016)\citenamefont{Tucci, Desjacques, and
  Kunz}}]{Tucci:2016hng}
\bibinfo{author}{\bibfnamefont{M.}~\bibnamefont{Tucci}},
  \bibinfo{author}{\bibfnamefont{V.}~\bibnamefont{Desjacques}},
  \bibnamefont{and} \bibinfo{author}{\bibfnamefont{M.}~\bibnamefont{Kunz}},
  \bibinfo{journal}{Mon. Not. Roy. Astron. Soc.}
  \textbf{\bibinfo{volume}{463}}, \bibinfo{pages}{2046} (\bibinfo{year}{2016}),
  \eprint{1606.02323}.

\bibitem[{\citenamefont{Lenz et~al.}(2019)\citenamefont{Lenz, Dor, and
  Lagache}}]{Lenz}
\bibinfo{author}{\bibfnamefont{D.}~\bibnamefont{Lenz}},
  \bibinfo{author}{\bibfnamefont{O.}~\bibnamefont{Dor}}, \bibnamefont{and}
  \bibinfo{author}{\bibfnamefont{G.}~\bibnamefont{Lagache}}
  (\bibinfo{year}{2019}), \eprint{1905.00426}.

\bibitem[{\citenamefont{Allen and Romano}(1999)}]{Allen:1997ad}
\bibinfo{author}{\bibfnamefont{B.}~\bibnamefont{Allen}} \bibnamefont{and}
  \bibinfo{author}{\bibfnamefont{J.~D.} \bibnamefont{Romano}},
  \bibinfo{journal}{Phys. Rev.} \textbf{\bibinfo{volume}{D59}},
  \bibinfo{pages}{102001} (\bibinfo{year}{1999}), \eprint{gr-qc/9710117}.

\bibitem[{\citenamefont{Maggiore}(2007)}]{Maggiore:1900zz}
\bibinfo{author}{\bibfnamefont{M.}~\bibnamefont{Maggiore}},
  \emph{\bibinfo{title}{{Gravitational Waves. Vol. 1: Theory and
  Experiments}}}, Oxford Master Series in Physics (\bibinfo{publisher}{Oxford
  University Press}, \bibinfo{year}{2007}), ISBN \bibinfo{isbn}{9780198570745,
  9780198520740},
  \urlprefix\url{http://www.oup.com/uk/catalogue/?ci=9780198570745}.

\bibitem[{\citenamefont{Arnowitt et~al.}(1961)\citenamefont{Arnowitt, Deser,
  and Misner}}]{Arnowitt:1961zz}
\bibinfo{author}{\bibfnamefont{R.~L.} \bibnamefont{Arnowitt}},
  \bibinfo{author}{\bibfnamefont{S.}~\bibnamefont{Deser}}, \bibnamefont{and}
  \bibinfo{author}{\bibfnamefont{C.~W.} \bibnamefont{Misner}},
  \bibinfo{journal}{Phys. Rev.} \textbf{\bibinfo{volume}{122}},
  \bibinfo{pages}{997} (\bibinfo{year}{1961}).

\bibitem[{\citenamefont{Flanagan and Hughes}(1998)}]{Flanagan:1997sx}
\bibinfo{author}{\bibfnamefont{E.~E.} \bibnamefont{Flanagan}} \bibnamefont{and}
  \bibinfo{author}{\bibfnamefont{S.~A.} \bibnamefont{Hughes}},
  \bibinfo{journal}{Phys. Rev.} \textbf{\bibinfo{volume}{D57}},
  \bibinfo{pages}{4535} (\bibinfo{year}{1998}), \eprint{gr-qc/9701039}.

\bibitem[{\citenamefont{Ferrari et~al.}(1999)\citenamefont{Ferrari, Matarrese,
  and Schneider}}]{Ferrari:1998jf}
\bibinfo{author}{\bibfnamefont{V.}~\bibnamefont{Ferrari}},
  \bibinfo{author}{\bibfnamefont{S.}~\bibnamefont{Matarrese}},
  \bibnamefont{and}
  \bibinfo{author}{\bibfnamefont{R.}~\bibnamefont{Schneider}},
  \bibinfo{journal}{Mon. Not. Roy. Astron. Soc.}
  \textbf{\bibinfo{volume}{303}}, \bibinfo{pages}{258} (\bibinfo{year}{1999}),
  \eprint{astro-ph/9806357}.

\bibitem[{\citenamefont{Schneider et~al.}(1998)\citenamefont{Schneider,
  Ferrari, and Matarrese}}]{Schneider:1998xt}
\bibinfo{author}{\bibfnamefont{R.}~\bibnamefont{Schneider}},
  \bibinfo{author}{\bibfnamefont{V.}~\bibnamefont{Ferrari}}, \bibnamefont{and}
  \bibinfo{author}{\bibfnamefont{S.}~\bibnamefont{Matarrese}}
  (\bibinfo{year}{1998}), \bibinfo{note}{[Nucl. Phys. Proc.
  Suppl.80,0722(2000)]}, \eprint{astro-ph/9903470}.

\bibitem[{\citenamefont{Schneider et~al.}(2001)\citenamefont{Schneider,
  Ferrari, Matarrese, and Portegies~Zwart}}]{Schneider:2000sg}
\bibinfo{author}{\bibfnamefont{R.}~\bibnamefont{Schneider}},
  \bibinfo{author}{\bibfnamefont{V.}~\bibnamefont{Ferrari}},
  \bibinfo{author}{\bibfnamefont{S.}~\bibnamefont{Matarrese}},
  \bibnamefont{and} \bibinfo{author}{\bibfnamefont{S.~F.}
  \bibnamefont{Portegies~Zwart}}, \bibinfo{journal}{Mon. Not. Roy. Astron.
  Soc.} \textbf{\bibinfo{volume}{324}}, \bibinfo{pages}{797}
  (\bibinfo{year}{2001}), \eprint{astro-ph/0002055}.

\bibitem[{\citenamefont{Abbott et~al.}(2016)}]{TheLIGOScientific:2016wyq}
\bibinfo{author}{\bibfnamefont{B.~P.} \bibnamefont{Abbott}}
  \bibnamefont{et~al.} (\bibinfo{collaboration}{LIGO Scientific, Virgo}),
  \bibinfo{journal}{Phys. Rev. Lett.} \textbf{\bibinfo{volume}{116}},
  \bibinfo{pages}{131102} (\bibinfo{year}{2016}), \eprint{1602.03847}.

\bibitem[{\citenamefont{Phinney}(2001)}]{Phinney:2001di}
\bibinfo{author}{\bibfnamefont{E.~S.} \bibnamefont{Phinney}}
  (\bibinfo{year}{2001}), \eprint{astro-ph/0108028}.

\bibitem[{\citenamefont{Nakazato et~al.}(2016)\citenamefont{Nakazato, Niino,
  and Sago}}]{Nakazato:2016nkj}
\bibinfo{author}{\bibfnamefont{K.}~\bibnamefont{Nakazato}},
  \bibinfo{author}{\bibfnamefont{Y.}~\bibnamefont{Niino}}, \bibnamefont{and}
  \bibinfo{author}{\bibfnamefont{N.}~\bibnamefont{Sago}},
  \bibinfo{journal}{Astrophys. J.} \textbf{\bibinfo{volume}{832}},
  \bibinfo{pages}{146} (\bibinfo{year}{2016}), \eprint{1605.02146}.

\bibitem[{\citenamefont{Maiolino et~al.}(2008)}]{Maiolino:2008gh}
\bibinfo{author}{\bibfnamefont{R.}~\bibnamefont{Maiolino}}
  \bibnamefont{et~al.}, \bibinfo{journal}{Astron. Astrophys.}
  \textbf{\bibinfo{volume}{488}}, \bibinfo{pages}{463} (\bibinfo{year}{2008}),
  \eprint{0806.2410}.

\bibitem[{\citenamefont{Behroozi et~al.}(2010)\citenamefont{Behroozi, Conroy,
  and Wechsler}}]{Behroozi:2010rx}
\bibinfo{author}{\bibfnamefont{P.~S.} \bibnamefont{Behroozi}},
  \bibinfo{author}{\bibfnamefont{C.}~\bibnamefont{Conroy}}, \bibnamefont{and}
  \bibinfo{author}{\bibfnamefont{R.~H.} \bibnamefont{Wechsler}},
  \bibinfo{journal}{Astrophys. J.} \textbf{\bibinfo{volume}{717}},
  \bibinfo{pages}{379} (\bibinfo{year}{2010}), \eprint{1001.0015}.

\bibitem[{\citenamefont{Behroozi et~al.}(2013)\citenamefont{Behroozi, Wechsler,
  and Conroy}}]{Behroozi:2012iw}
\bibinfo{author}{\bibfnamefont{P.~S.} \bibnamefont{Behroozi}},
  \bibinfo{author}{\bibfnamefont{R.~H.} \bibnamefont{Wechsler}},
  \bibnamefont{and} \bibinfo{author}{\bibfnamefont{C.}~\bibnamefont{Conroy}},
  \bibinfo{journal}{Astrophys. J.} \textbf{\bibinfo{volume}{770}},
  \bibinfo{pages}{57} (\bibinfo{year}{2013}), \eprint{1207.6105}.

\bibitem[{\citenamefont{Press and Schechter}(1974)}]{Press:1973iz}
\bibinfo{author}{\bibfnamefont{W.~H.} \bibnamefont{Press}} \bibnamefont{and}
  \bibinfo{author}{\bibfnamefont{P.}~\bibnamefont{Schechter}},
  \bibinfo{journal}{Astrophys. J.} \textbf{\bibinfo{volume}{187}},
  \bibinfo{pages}{425} (\bibinfo{year}{1974}).

\bibitem[{\citenamefont{Sheth and Tormen}(1999)}]{Sheth:1999mn}
\bibinfo{author}{\bibfnamefont{R.~K.} \bibnamefont{Sheth}} \bibnamefont{and}
  \bibinfo{author}{\bibfnamefont{G.}~\bibnamefont{Tormen}},
  \bibinfo{journal}{Mon. Not. Roy. Astron. Soc.}
  \textbf{\bibinfo{volume}{308}}, \bibinfo{pages}{119} (\bibinfo{year}{1999}),
  \eprint{astro-ph/9901122}.

\bibitem[{\citenamefont{Tinker et~al.}(2008{\natexlab{a}})\citenamefont{Tinker,
  Kravtsov, Klypin, Abazajian, Warren, Yepes, Gottlober, and
  Holz}}]{Tinker:2008ff}
\bibinfo{author}{\bibfnamefont{J.~L.} \bibnamefont{Tinker}},
  \bibinfo{author}{\bibfnamefont{A.~V.} \bibnamefont{Kravtsov}},
  \bibinfo{author}{\bibfnamefont{A.}~\bibnamefont{Klypin}},
  \bibinfo{author}{\bibfnamefont{K.}~\bibnamefont{Abazajian}},
  \bibinfo{author}{\bibfnamefont{M.~S.} \bibnamefont{Warren}},
  \bibinfo{author}{\bibfnamefont{G.}~\bibnamefont{Yepes}},
  \bibinfo{author}{\bibfnamefont{S.}~\bibnamefont{Gottlober}},
  \bibnamefont{and} \bibinfo{author}{\bibfnamefont{D.~E.} \bibnamefont{Holz}},
  \bibinfo{journal}{Astrophys. J.} \textbf{\bibinfo{volume}{688}},
  \bibinfo{pages}{709} (\bibinfo{year}{2008}{\natexlab{a}}),
  \eprint{0803.2706}.

\bibitem[{\citenamefont{Springel and Hernquist}(2003)}]{Springel:2002ux}
\bibinfo{author}{\bibfnamefont{V.}~\bibnamefont{Springel}} \bibnamefont{and}
  \bibinfo{author}{\bibfnamefont{L.}~\bibnamefont{Hernquist}},
  \bibinfo{journal}{Mon. Not. Roy. Astron. Soc.}
  \textbf{\bibinfo{volume}{339}}, \bibinfo{pages}{312} (\bibinfo{year}{2003}),
  \eprint{astro-ph/0206395}.

\bibitem[{\citenamefont{Hernquist and Springel}(2003)}]{Hernquist:2002rg}
\bibinfo{author}{\bibfnamefont{L.}~\bibnamefont{Hernquist}} \bibnamefont{and}
  \bibinfo{author}{\bibfnamefont{V.}~\bibnamefont{Springel}},
  \bibinfo{journal}{Mon. Not. Roy. Astron. Soc.}
  \textbf{\bibinfo{volume}{341}}, \bibinfo{pages}{1253} (\bibinfo{year}{2003}),
  \eprint{astro-ph/0209183}.

\bibitem[{\citenamefont{Shapiro}(1964)}]{PhysRevLett.13.789}
\bibinfo{author}{\bibfnamefont{I.~I.} \bibnamefont{Shapiro}},
  \bibinfo{journal}{Phys. Rev. Lett.} \textbf{\bibinfo{volume}{13}},
  \bibinfo{pages}{789} (\bibinfo{year}{1964}),
  \urlprefix\url{https://link.aps.org/doi/10.1103/PhysRevLett.13.789}.

\bibitem[{\citenamefont{Fanizza et~al.}(2018)\citenamefont{Fanizza, Yoo, and
  Biern}}]{Fanizza:2018qux}
\bibinfo{author}{\bibfnamefont{G.}~\bibnamefont{Fanizza}},
  \bibinfo{author}{\bibfnamefont{J.}~\bibnamefont{Yoo}}, \bibnamefont{and}
  \bibinfo{author}{\bibfnamefont{S.~G.} \bibnamefont{Biern}},
  \bibinfo{journal}{JCAP} \textbf{\bibinfo{volume}{1809}}, \bibinfo{pages}{037}
  (\bibinfo{year}{2018}), \eprint{1805.05959}.

\bibitem[{\citenamefont{Biern and Yoo}(2017)}]{Biern:2016kys}
\bibinfo{author}{\bibfnamefont{S.~G.} \bibnamefont{Biern}} \bibnamefont{and}
  \bibinfo{author}{\bibfnamefont{J.}~\bibnamefont{Yoo}},
  \bibinfo{journal}{JCAP} \textbf{\bibinfo{volume}{1704}}, \bibinfo{pages}{045}
  (\bibinfo{year}{2017}), \eprint{1606.01910}.

\bibitem[{\citenamefont{Bellomo et~al.}(2019)}]{SGWB-II}
\bibinfo{author}{\bibnamefont{Bellomo}} \bibnamefont{et~al.},
  \bibinfo{journal}{In Prep.}  (\bibinfo{year}{2019}).

\bibitem[{\citenamefont{Regimbau et~al.}(2017)\citenamefont{Regimbau, Evans,
  Christensen, Katsavounidis, Sathyaprakash, and Vitale}}]{Regimbau:2016ike}
\bibinfo{author}{\bibfnamefont{T.}~\bibnamefont{Regimbau}},
  \bibinfo{author}{\bibfnamefont{M.}~\bibnamefont{Evans}},
  \bibinfo{author}{\bibfnamefont{N.}~\bibnamefont{Christensen}},
  \bibinfo{author}{\bibfnamefont{E.}~\bibnamefont{Katsavounidis}},
  \bibinfo{author}{\bibfnamefont{B.}~\bibnamefont{Sathyaprakash}},
  \bibnamefont{and} \bibinfo{author}{\bibfnamefont{S.}~\bibnamefont{Vitale}},
  \bibinfo{journal}{Phys. Rev. Lett.} \textbf{\bibinfo{volume}{118}},
  \bibinfo{pages}{151105} (\bibinfo{year}{2017}), \eprint{1611.08943}.

\bibitem[{\citenamefont{Ajith et~al.}(2011)\citenamefont{Ajith, Hannam, Husa,
  Chen, Br\"ugmann, Dorband, M\"uller, Ohme, Pollney, Reisswig
  et~al.}}]{ajith:gwswaveform}
\bibinfo{author}{\bibfnamefont{P.}~\bibnamefont{Ajith}},
  \bibinfo{author}{\bibfnamefont{M.}~\bibnamefont{Hannam}},
  \bibinfo{author}{\bibfnamefont{S.}~\bibnamefont{Husa}},
  \bibinfo{author}{\bibfnamefont{Y.}~\bibnamefont{Chen}},
  \bibinfo{author}{\bibfnamefont{B.}~\bibnamefont{Br\"ugmann}},
  \bibinfo{author}{\bibfnamefont{N.}~\bibnamefont{Dorband}},
  \bibinfo{author}{\bibfnamefont{D.}~\bibnamefont{M\"uller}},
  \bibinfo{author}{\bibfnamefont{F.}~\bibnamefont{Ohme}},
  \bibinfo{author}{\bibfnamefont{D.}~\bibnamefont{Pollney}},
  \bibinfo{author}{\bibfnamefont{C.}~\bibnamefont{Reisswig}},
  \bibnamefont{et~al.}, \bibinfo{journal}{Phys. Rev. Lett.}
  \textbf{\bibinfo{volume}{106}}, \bibinfo{pages}{241101}
  (\bibinfo{year}{2011}), \eprint{0909.2867}.

\bibitem[{\citenamefont{Tinker et~al.}(2010)\citenamefont{Tinker, Robertson,
  Kravtsov, Klypin, Warren, Yepes, and Gottl{\"o}ber}}]{tinker:halobias}
\bibinfo{author}{\bibfnamefont{J.~L.} \bibnamefont{Tinker}},
  \bibinfo{author}{\bibfnamefont{B.~E.} \bibnamefont{Robertson}},
  \bibinfo{author}{\bibfnamefont{A.~V.} \bibnamefont{Kravtsov}},
  \bibinfo{author}{\bibfnamefont{A.}~\bibnamefont{Klypin}},
  \bibinfo{author}{\bibfnamefont{M.~S.} \bibnamefont{Warren}},
  \bibinfo{author}{\bibfnamefont{G.}~\bibnamefont{Yepes}}, \bibnamefont{and}
  \bibinfo{author}{\bibfnamefont{S.}~\bibnamefont{Gottl{\"o}ber}},
  \bibinfo{journal}{The Astrophysical Journal} \textbf{\bibinfo{volume}{724}},
  \bibinfo{pages}{878} (\bibinfo{year}{2010}), \eprint{1104.3565}.

\bibitem[{\citenamefont{Tinker et~al.}(2008{\natexlab{b}})\citenamefont{Tinker,
  Kravtsov, Klypin, Abazajian, Warren, Yepes, Gottl{\"o}ber, and
  Holz}}]{tinker:halomassfunction}
\bibinfo{author}{\bibfnamefont{J.}~\bibnamefont{Tinker}},
  \bibinfo{author}{\bibfnamefont{A.~V.} \bibnamefont{Kravtsov}},
  \bibinfo{author}{\bibfnamefont{A.}~\bibnamefont{Klypin}},
  \bibinfo{author}{\bibfnamefont{K.}~\bibnamefont{Abazajian}},
  \bibinfo{author}{\bibfnamefont{M.}~\bibnamefont{Warren}},
  \bibinfo{author}{\bibfnamefont{G.}~\bibnamefont{Yepes}},
  \bibinfo{author}{\bibfnamefont{S.}~\bibnamefont{Gottl{\"o}ber}},
  \bibnamefont{and} \bibinfo{author}{\bibfnamefont{D.~E.} \bibnamefont{Holz}},
  \bibinfo{journal}{The Astrophysical Journal} \textbf{\bibinfo{volume}{688}},
  \bibinfo{pages}{709} (\bibinfo{year}{2008}{\natexlab{b}}),
  \eprint{0803.2706}.

\bibitem[{\citenamefont{Hall et~al.}(2013)\citenamefont{Hall, Bonvin, and
  Challinor}}]{Hall:2012wd}
\bibinfo{author}{\bibfnamefont{A.}~\bibnamefont{Hall}},
  \bibinfo{author}{\bibfnamefont{C.}~\bibnamefont{Bonvin}}, \bibnamefont{and}
  \bibinfo{author}{\bibfnamefont{A.}~\bibnamefont{Challinor}},
  \bibinfo{journal}{Phys. Rev.} \textbf{\bibinfo{volume}{D87}},
  \bibinfo{pages}{064026} (\bibinfo{year}{2013}), \eprint{1212.0728}.

\bibitem[{\citenamefont{Alonso and Ferreira}(2015)}]{Alonso:2015sfa}
\bibinfo{author}{\bibfnamefont{D.}~\bibnamefont{Alonso}} \bibnamefont{and}
  \bibinfo{author}{\bibfnamefont{P.~G.} \bibnamefont{Ferreira}},
  \bibinfo{journal}{Phys. Rev.} \textbf{\bibinfo{volume}{D92}},
  \bibinfo{pages}{063525} (\bibinfo{year}{2015}), \eprint{1507.03550}.

\end{thebibliography}

\end{document}